\pdfoutput=1
\documentclass[11pt,a4paper]{article}
\usepackage{jheppub}
\usepackage{microtype}
\usepackage{dcolumn}
\usepackage{booktabs}

\usepackage{lineno}  

\usepackage{xspace}

\def\bfseries{\fontseries \bfdefault \selectfont \boldmath}

\newcommand{\Ba}[1]{\ensuremath{^{#1}\mathrm{Ba}}\xspace}
\newcommand{\Bi}[1]{\ensuremath{^{#1}\mathrm{Bi}}\xspace}
\newcommand{\Co}[1]{\ensuremath{^{#1}\mathrm{Co}}\xspace}
\newcommand{\Cs}[1]{\ensuremath{^{#1}\mathrm{Cs}}\xspace}
\newcommand{\K}[1]{\ensuremath{^{#1}\mathrm{K}}\xspace}
\newcommand{\Kr}[1]{\ensuremath{^{#1}\mathrm{Kr}}\xspace}

\newcommand{\Rb}[1]{\ensuremath{^{#1}\mathrm{Rb}}\xspace}
\newcommand{\Rn}[1]{\ensuremath{^{#1}\mathrm{Rn}}\xspace}
\newcommand{\Th}[1]{\ensuremath{^{#1}\mathrm{Th}}\xspace}
\newcommand{\Tl}[1]{\ensuremath{^{#1}\mathrm{Tl}}\xspace}

\newcommand{\Xe}[1]{\ensuremath{^{#1}\mathrm{Xe}}\xspace}

\newcommand{\bbnonu}{\ensuremath{0\nu\beta\beta}\xspace}
\newcommand{\bbtwonu}{\ensuremath{2\nu\beta\beta}\xspace}
\newcommand{\bb}{\ensuremath{\beta\beta}\xspace}
\newcommand{\Qbb}{\ensuremath{Q_{\beta\beta}}\xspace}

\newcommand{\halflife}{\ensuremath{T_{1/2}}\xspace}

\begin{document}

\title{Radiogenic backgrounds in the NEXT double beta decay experiment}

\collaboration{The NEXT Collaboration}
\author[19,a]{P.~Novella,\note[a]{Corresponding author.}}
\author[19]{B.~Palmeiro,}
\author[19]{M.~Sorel,}
\author[19]{A.~Us\'on,}
\author[16,9]{P.~Ferrario,}
\author[16,9,b]{J.J.~G\'omez-Cadenas,\note[b]{NEXT Co-spokesperson.}}
\author[11]{C.~Adams,}
\author[19]{V.~\'Alvarez,}
\author[6]{L.~Arazi,}
\author[20]{I.J.~Arnquist,}
\author[4]{C.D.R~Azevedo,}
\author[2]{K.~Bailey,}
\author[22]{F.~Ballester,}
\author[19]{J.M.~Benlloch-Rodr\'{i}guez,}
\author[14]{F.I.G.M.~Borges,}
\author[3]{N.~Byrnes,}
\author[19]{S.~C\'arcel,}
\author[19]{J.V.~Carri\'on,}
\author[23]{S.~Cebri\'an,}
\author[20]{E.~Church,}
\author[14]{C.A.N.~Conde,}
\author[11]{T.~Contreras,}
\author[21,16]{G.~D\'iaz,}
\author[19]{J.~D\'iaz,}
\author[5]{M.~Diesburg,}
\author[14]{J.~Escada,}
\author[22]{R.~Esteve,}
\author[19]{R.~Felkai,}
\author[13]{A.F.M.~Fernandes,}
\author[13]{L.M.P.~Fernandes,}
\author[4]{A.L.~Ferreira,}
\author[13]{E.D.C.~Freitas,}
\author[16]{J.~Generowicz,}
\author[11]{S.~Ghosh,}
\author[8]{A.~Goldschmidt,}
\author[21]{D.~Gonz\'alez-D\'iaz,}
\author[11]{R.~Guenette,}
\author[10]{R.M.~Guti\'errez,}
\author[11]{J.~Haefner,}
\author[2]{K.~Hafidi,}
\author[1]{J.~Hauptman,}
\author[13]{C.A.O.~Henriques,}
\author[21]{J.A.~Hernando~Morata,}
\author[16,19]{P.~Herrero,}
\author[22]{V.~Herrero,}
\author[6,7]{Y.~Ifergan,}
\author[2]{S.~Johnston,}
\author[3]{B.J.P.~Jones,}
\author[19]{M.~Kekic,}
\author[18]{L.~Labarga,}
\author[3]{A.~Laing,}
\author[5]{P.~Lebrun,}
\author[19]{N.~L\'opez-March,}
\author[10]{M.~Losada,}
\author[13]{R.D.P.~Mano,}
\author[11]{J.~Mart\'in-Albo,}
\author[16]{A.~Mart\'inez,}
\author[19,21,c]{G.~Mart\'inez-Lema,\note[c]{Now at Weizmann Institute of Science, Israel.}}
\author[3]{A.D.~McDonald,}
\author[16,9]{F.~Monrabal,}
\author[13]{C.M.B.~Monteiro,}
\author[22]{F.J.~Mora,}
\author[19]{J.~Mu\~noz Vidal,}
\author[3,d]{D.R.~Nygren,\note[d]{NEXT Co-spokesperson.}}
\author[5]{A.~Para,}
\author[12]{J.~P\'erez,}
\author[3]{F.~Psihas,}
\author[19]{M.~Querol,}
\author[19]{J.~Renner,}
\author[2]{J.~Repond,}
\author[2]{S.~Riordan,}
\author[17]{L.~Ripoll,}
\author[10]{Y.~Rodr\'iguez Garc\'ia,}
\author[22]{J.~Rodr\'iguez,}
\author[3]{L.~Rogers,}
\author[16,12]{B.~Romeo,}
\author[19]{C.~Romo-Luque,}
\author[14]{F.P.~Santos,}
\author[13]{J.M.F. dos~Santos,}
\author[6]{A.~Sim\'on,}
\author[15,e]{C.~Sofka,\note[e]{Now at University of Texas at Austin, USA.}}
\author[15]{T.~Stiegler,}
\author[22]{J.F.~Toledo,}
\author[16]{J.~Torrent,}
\author[4]{J.F.C.A.~Veloso,}
\author[15]{R.~Webb,}
\author[6,f]{R.~Weiss-Babai,\note[f]{On leave from Soreq Nuclear Research Center, Yavneh, Israel.}}
\author[15,g]{J.T.~White,\note[g]{Deceased.}}
\author[3]{K.~Woodruff,}
\author[19]{N.~Yahlali}
\emailAdd{pau.novella@ific.uv.es}
\affiliation[1]{
Department of Physics and Astronomy, Iowa State University, 12 Physics Hall, Ames, IA 50011-3160, USA}
\affiliation[2]{
Argonne National Laboratory, Argonne, IL 60439, USA}
\affiliation[3]{
Department of Physics, University of Texas at Arlington, Arlington, TX 76019, USA}
\affiliation[4]{
Institute of Nanostructures, Nanomodelling and Nanofabrication (i3N), Universidade de Aveiro, Campus de Santiago, Aveiro, 3810-193, Portugal}
\affiliation[5]{
Fermi National Accelerator Laboratory, Batavia, IL 60510, USA}
\affiliation[6]{
Nuclear Engineering Unit, Faculty of Engineering Sciences, Ben-Gurion University of the Negev, P.O.B. 653, Beer-Sheva, 8410501, Israel}
\affiliation[7]{
Nuclear Research Center Negev, Beer-Sheva, 84190, Israel}
\affiliation[8]{
Lawrence Berkeley National Laboratory (LBNL), 1 Cyclotron Road, Berkeley, CA 94720, USA}
\affiliation[9]{
Ikerbasque, Basque Foundation for Science, Bilbao, E-48013, Spain}
\affiliation[10]{
Centro de Investigaci\'on en Ciencias B\'asicas y Aplicadas, Universidad Antonio Nari\~no, Sede Circunvalar, Carretera 3 Este No.\ 47 A-15, Bogot\'a, Colombia}
\affiliation[11]{
Department of Physics, Harvard University, Cambridge, MA 02138, USA}
\affiliation[12]{
Laboratorio Subterr\'aneo de Canfranc, Paseo de los Ayerbe s/n, Canfranc Estaci\'on, E-22880, Spain}
\affiliation[13]{
LIBPhys, Physics Department, University of Coimbra, Rua Larga, Coimbra, 3004-516, Portugal}
\affiliation[14]{
LIP, Department of Physics, University of Coimbra, Coimbra, 3004-516, Portugal}
\affiliation[15]{
Department of Physics and Astronomy, Texas A\&M University, College Station, TX 77843-4242, USA}
\affiliation[16]{
Donostia International Physics Center (DIPC), Paseo Manuel Lardizabal, 4, Donostia-San Sebastian, E-20018, Spain}
\affiliation[17]{
Escola Polit\`ecnica Superior, Universitat de Girona, Av.~Montilivi, s/n, Girona, E-17071, Spain}
\affiliation[18]{
Departamento de F\'isica Te\'orica, Universidad Aut\'onoma de Madrid, Campus de Cantoblanco, Madrid, E-28049, Spain}
\affiliation[19]{
Instituto de F\'isica Corpuscular (IFIC), CSIC \& Universitat de Val\`encia, Calle Catedr\'atico Jos\'e Beltr\'an, 2, Paterna, E-46980, Spain}
\affiliation[20]{
Pacific Northwest National Laboratory (PNNL), Richland, WA 99352, USA}
\affiliation[21]{
Instituto Gallego de F\'isica de Altas Energ\'ias, Univ.\ de Santiago de Compostela, Campus sur, R\'ua Xos\'e Mar\'ia Su\'arez N\'u\~nez, s/n, Santiago de Compostela, E-15782, Spain}
\affiliation[22]{
Instituto de Instrumentaci\'on para Imagen Molecular (I3M), Centro Mixto CSIC - Universitat Polit\`ecnica de Val\`encia, Camino de Vera s/n, Valencia, E-46022, Spain}
\affiliation[23]{
Laboratorio de F\'isica Nuclear y Astropart\'iculas, Universidad de Zaragoza, Calle Pedro Cerbuna, 12, Zaragoza, E-50009, Spain}

\abstract{ Natural radioactivity represents one of the main backgrounds in the search for neutrinoless double beta decay. Within the NEXT physics program, the radioactivity-induced backgrounds are measured with the NEXT-White detector. Data from 37.9~days of low-background operations at the Laboratorio Subterr\'aneo de Canfranc with xenon depleted in $^{136}$Xe are analyzed to derive a total background rate of (0.84$\pm$0.02)~mHz above 1000~keV. The comparison of data samples with and without the use of the radon abatement system demonstrates that the contribution of airborne-Rn is negligible. A radiogenic background model is built upon the extensive radiopurity screening campaign conducted by the NEXT Collaboration. A spectral fit to this model yields the specific contributions of \Co{60}, \K{40}, \Bi{214} and \Tl{208} to the total background rate, as well as their location in the detector volumes. The results are used to evaluate the impact of the radiogenic backgrounds in the double beta decay analyses, after the application of topological cuts that reduce the total rate to (0.25$\pm$0.01)~mHz. Based on the best-fit background model, the NEXT-White median sensitivity to the two-neutrino double beta decay is found to be 3.5$\sigma$ after 1 year of data taking. The background measurement in a Q$_{\bb}\pm$100 keV energy window validates the best-fit background model also for the neutrinoless double beta decay search with NEXT-100. Only one event is found, while the model expectation is (0.75$\pm$0.12) events.}

\maketitle
       
\clearpage

\section{Introduction}
\label{sec:intro}


The results from oscillation experiments in the last decades have demonstrated that neutrinos are massive particles and that lepton flavor is not conserved. As a consequence, the double beta (\bb) decay experiments play nowadays a major role in understanding the nature of the neutrino masses. The \bb decay is a nuclear transition in which two neutrons bound in a nucleus  are  simultaneously transformed into two protons plus two electrons. Although highly suppressed, this transition can occur for nuclei in which the $\beta$-decay is highly forbidden or energetically not allowed. The decay mode in which two neutrinos are emitted (\bbtwonu) has been observed in many nuclei, with typical half-lives in the $10^{19}-10^{21}$~yr range. However, the neutrinoless double beta (\bbnonu) decay, violating lepton number conservation, has not been detected. Regardless of the underlying decay mechanism, the observation of this process would demonstrate the Majorana nature of neutrinos.    


According to the current best limits \cite{KamLAND-Zen:2016pfg,Agostini:2018tnm}, the half-life of the \bbnonu decay is above $\sim$10$^{26}$~yr. This implies a significant experimental challenge which is being addressed by developing detector technologies that offer at the same time good energy resolutions and background rejection capabilities. In addition, any detector must rely on very radiopure materials. Given the relatively low Q-values of the \bb-emitter isotopes (Q$_{\bb}$), natural radioactivity of detector materials becomes one of the main backgrounds in the search for the \bbnonu decay. The total background budget is completed with contributions from \bbtwonu events, airborne-radon ($\beta$ decays of the \Rn{220} and \Rn{222} progeny) and cosmogenic events (prompt gammas following n-captures and radioactive nuclei activation). The evaluation and characterization of the different background sources are key elements in the data analysis, as the identification of the \bbnonu signal is based on the excess of events in a given energy window. 


Within the physics program of the Neutrino Experiment with a Xenon Time Projection Chamber (NEXT), the measurement of the \bb backgrounds is one of the major goals of the NEXT-White detector, currently operating at the Laboratorio Subterr\'aneo de Canfranc (LSC).  The detector technology exploited by the NEXT collaboration to search for the \bbnonu decay is a high-pressure (10--15 bar) \Xe{136} gas time projection chamber (TPC) \cite{Alvarez:2012flf}. Xenon is the only noble gas that has a \bb-decaying isotope and no other long-lived radioactive isotope. Its \Qbb\ value is relatively high (\Qbb= 2457.83$\pm$0.37 ~keV \cite{Redshaw:2007un}) and the half-life of the \bbtwonu mode has been measured to be in excess of 10$^{21}$~yr \cite{Albert:2013gpz,KamLANDZen:2012aa}, thus being a suitable isotope as far as backgrounds are concerned. A xenon TPC provides both primary scintillation light and ionization electrons when charged particles pass through the active volume. The scintillation light is used to determine the start time of the event, while the ionization electrons provide a measurement of the event energy and topology. The ionization signal amplification by means of electroluminescence (secondary scintillation light) allows for a demonstrated energy resolution of 1\% FWHM at the \Qbb of \Xe{136} \cite{Renner:2018ttw,Renner:2019pfe}, which can be improved according to results at lower energies \cite{Alvarez:2012yxw,Lorca:2014sra}. In addition, the detector low-density and fine spatial granularity of the tracking readout provides an efficient identification of the topological signature characteristic of \bbnonu \cite{Ferrario:2015kta,Ferrario:2019kwg}. Finally, this technology offers promising \Ba{136} (daughter of \Xe{136}) tagging capabilities \cite{McDonald:2017izm,Thapa:2019zjk}. The implementation of an effective \Ba{136}-tagging would imply a background-free experiment.
 
After a successful R\&D phase in 2008--2014 \cite{Alvarez:2012hu,Alvarez:2012zsz,Alvarez:2013gxa,Alvarez:2013oha,Renner:2014mha,Serra:2014zda,Gonzalez-Diaz:2015oba}, the experiment has started underground operations at the LSC with the NEXT-White detector, holding about 5~kg of \Xe{} \cite{Monrabal:2018xlr}. While the operation with \Xe{136}-depleted xenon allows for the calibration of the detector and the \bb background characterization (as presented in this work), the operation with xenon enriched in \Xe{136} will allow for the measurement of the \bbtwonu half-life. The technology of NEXT-White is being scaled up in order to build the NEXT-100 detector at the LSC, using 100 kg of \Xe{136}. The sensitivity of NEXT-100 to the \bbnonu decay has been evaluated in \cite{Martin-Albo:2015rhw}, relying on detailed radio-assay measurements \cite{Alvarez:2012as,Alvarez:2014kvs,Cebrian:2017jzb} and Monte-Carlo simulations. While the assumptions concerning the internal Radon-induced backgrounds have been validated with the NEXT-White data in \cite{Novella:2018ewv}, this work presents a first measurement of the detector-induced backgrounds which validates the inclusive background model based on Monte-Carlo.     


This paper is organized as follows. Sec.~\ref{sec:nextwhite} gives a description of the NEXT-White detector, as well as the operating conditions and facilities at the LSC. The event reconstruction and fiducial selection are discussed in Sec. \ref{sec:fiducialsel}. Sec.~\ref{sec:backgrounddata} describes the different data taking periods during the so-called Run-IV and provides the corresponding background measurements. Sec.~\ref{sec:backgroundmodel} presents a comprehensive description of the Monte-Carlo background model, which is compared with Run-IV data in Sec.~\ref{sec:backgroundfit}. Finally, Sec.~\ref{sec:topology} presents the topological selection of double-electron events, while Sec.~\ref{sec:betabetasearch} estimates the corresponding background in the \bb analyses. 

\section{The NEXT-White detector}
\label{sec:nextwhite}


The technological approach adopted by the NEXT collaboration to search for the \bbnonu decay is a high-pressure gaseous xenon TPC \cite{Alvarez:2012flf,Monrabal:2018xlr}. A charged particle interacting in the active volume produces both primary scintillation light (S1) and ionization electrons. The ionization charge is drifted to the anode, where secondary scintillation (S2) is produced by means of the electroluminescence process (EL). This allows to measure both scintillation and ionization signals with the same photosensors, as well as to optimize the energy resolution. The S2 light is read by two planes of photo-detectors located at opposite ends of the detector cylindrical structure, allowing for both the energy and tracking measurements. The readout plane behind the transparent cathode (energy plane) performs the energy measurement by detecting the backward EL light using an array of low-radioactivity photomultipliers (PMTs). These sensors are also used to determine the initial time of the event ($t_0$) by collecting the S1 light. The energy plane is thereby used to trigger the detector using either the S1 or S2 light. The readout plane behind the anode (tracking plane), located a few mm away from the EL gap, provides the event topology by detecting the forward EL light with a dense array of silicon photomultipliers (SiPMs).

\begin{figure}
  \begin{center}
    \includegraphics[scale=0.4]{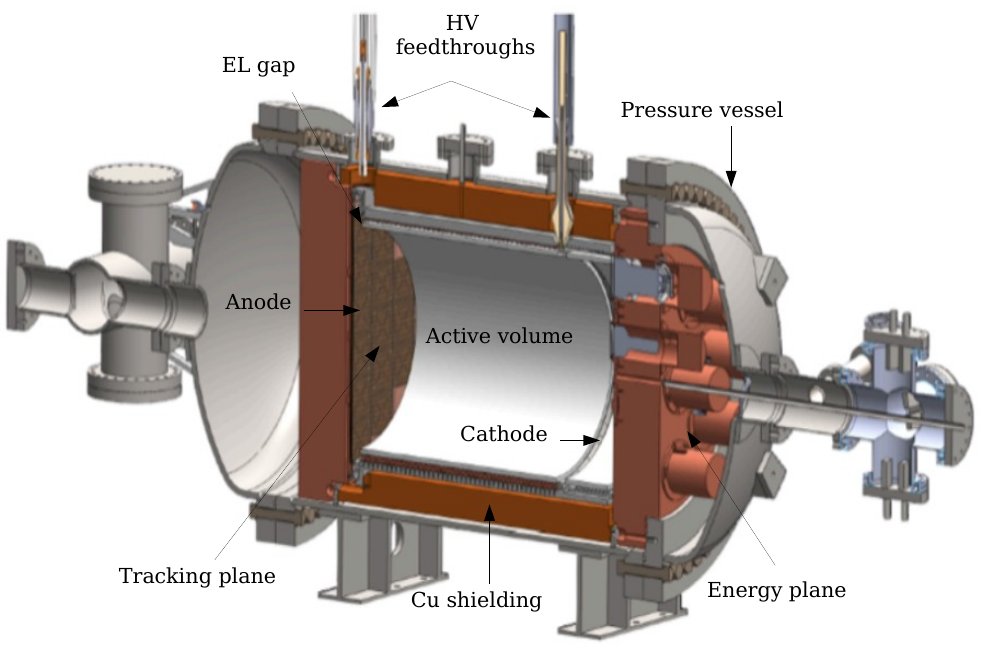}
    \caption{Schematic view of the NEXT-White detector and its main components.}
    \label{fig:NEW}
  \end{center}
\end{figure}

The NEXT-White detector\footnote{Named after Prof.~James White, our late mentor and friend.}, located at the LSC (Spain), is the first radiopure implementation of the NEXT TPC. A comprehensive description of NEXT-White can be found in \cite{Monrabal:2018xlr}. The active volume is 530.3~mm long along the drift direction, with a 198~mm radius. The energy plane read-out consists of 12 Hamamatsu R11410-10 PMTs. The tracking plane read-out consists of 1792 SensL C-Series SiPMs. In order to reduce the external backgrounds, a 6~cm thick copper shield within the pressure vessel has been installed. A schematic view of the detector is shown in Fig.\ref{fig:NEW}. The detector lies on a seismic platform and is surrounded by an additional 20~cm thick shield structure made of lead bricks (outer lead castle). In December 2018, a second lead structure (inner lead castle or ILC) has been installed to provide further shielding against external backgrounds. A radon abatement system (RAS) by ATEKO A.S. has been flushing radon-free air into the air volume enclosed by the lead castle since October 2018. The \Rn{222} content in the flushed air is 4--5 orders of magnitude lower compared to LSC Hall A air \cite{Novella:2018ewv}. As demonstrated in Sec.~\ref{sec:backgrounddata}, such a reduction allows the operation of NEXT-White (and in the future, NEXT-100) in a virtually airborne-Rn-free environment. The main scientific goals of NEXT-White are the technology certification for the NEXT-100 detector, the validation of the NEXT background model, and a measurement of the \Xe{136} \bbtwonu decay mode.

The detector was operated with \Xe{136}-depleted xenon ($\simeq$3\% isotopic abundance) between October 2016 and January 2019 (Run I--IV), and has been operating with \Xe{136}-enriched xenon ($\simeq$91\% isotopic abundance) since February 2019 (Run-V). After a short commissioning period (Run-I), the first calibration data-taking period took place from March 2017 to November 2017 (Run-II). At the end of Run-II, the field cage resistor chain and the PMT bases were replaced in order to improve their radiopurity. Following a short engineering run (Run-III), Run-IV lasted from July 2018 to January 2019, comprising a high-energy calibration campaign and a low-background data-taking period. The gas pressure, drift field and EL field were set to 10.1~bar, 0.4~kV/cm and 1.7~kV/(cm$\cdot$bar), respectively. For these conditions, the electron drift velocity was measured to be 0.92~mm/$\mu$s \cite{Simon:2018vep}, with sub-percent variations during Run-IV. The gas purity improved continuously with time as gas recirculated through a heated getter-based purifier MonoTorr PS4-MT50-R from SAES. However, a significant dependence on the temperature in the HALL A of the LSC was found. The lifetime ranged from $\sim$2000~$\mu$s at the beginning of Run-IV, to $\sim$5000~$\mu$s at the end. Continuous detector calibration and monitoring was carried out during Run-IV with a \Kr{83m} low-energy calibration source, ensuring high-quality and properly calibrated low-background data \cite{Martinez-Lema:2018ibw}. This was possible thanks to a dual-trigger implementation in the data acquisition system (DAQ) which allowed us to collect both low-energy ($\lesssim$100 keV) and high-energy ($\gtrsim$400 keV) events within the same DAQ run. A high-energy calibration campaign deploying \Cs{137} and \Th{232} calibration sources was performed during Run-IV. High-energy calibration data have been used to calibrate the detector energy scale (see Sec.~\ref{sec:backgrounddata}) and to validate event selection efficiencies (see Secs.~\ref{sec:fiducialsel}, \ref{sec:topology}) of low-background data. Low-background data runs (no calibration sources deployed other than \Kr{83m} and loose trigger conditions) were taken for about 5 months during Run-IV. The results presented in this work are based on these data. The background measurement as well as the study of its different contributions can be extrapolated to the ongoing Run-V, devoted to the measurement of the \Xe{136} \bbtwonu half-life, as the operating conditions (gas pressure, TPC voltages) are the same.

\section{Event reconstruction and fiducial selection}
\label{sec:fiducialsel}

Collected triggers are processed according to custom-made reconstruction algorithms. First, binary data are converted into PMT and SiPM waveforms, which are in turn serialized in a convenient data format for analysis. Second, the PMT waveforms are processed to zero-suppress the data and to find the S1 and S2 signals. Third, the SiPM hits providing the X and Y coordinates are reconstructed separately for each time (or Z) slice of the S2 signals. A veto against alpha particles based on the amplitude of the S1 signal is also applied to successfully reconstructed data events, see \cite{Novella:2018ewv}. Two basic selection procedures are then applied to data and MC reconstructed events. The so-called inclusive selection requires only one S2 signal per event. The fiducial selection requires in addition that no 3D hits are reconstructed within 20~mm from the detector boundaries. This cut reduces significantly the surface backgrounds, rejecting all $\beta$ particles entering the active volume. The remaining external backgrounds in the data sample are those induced by gammas interacting in the fiducial volume.

\begin{figure}
  \begin{center}
    \includegraphics[scale=0.4]{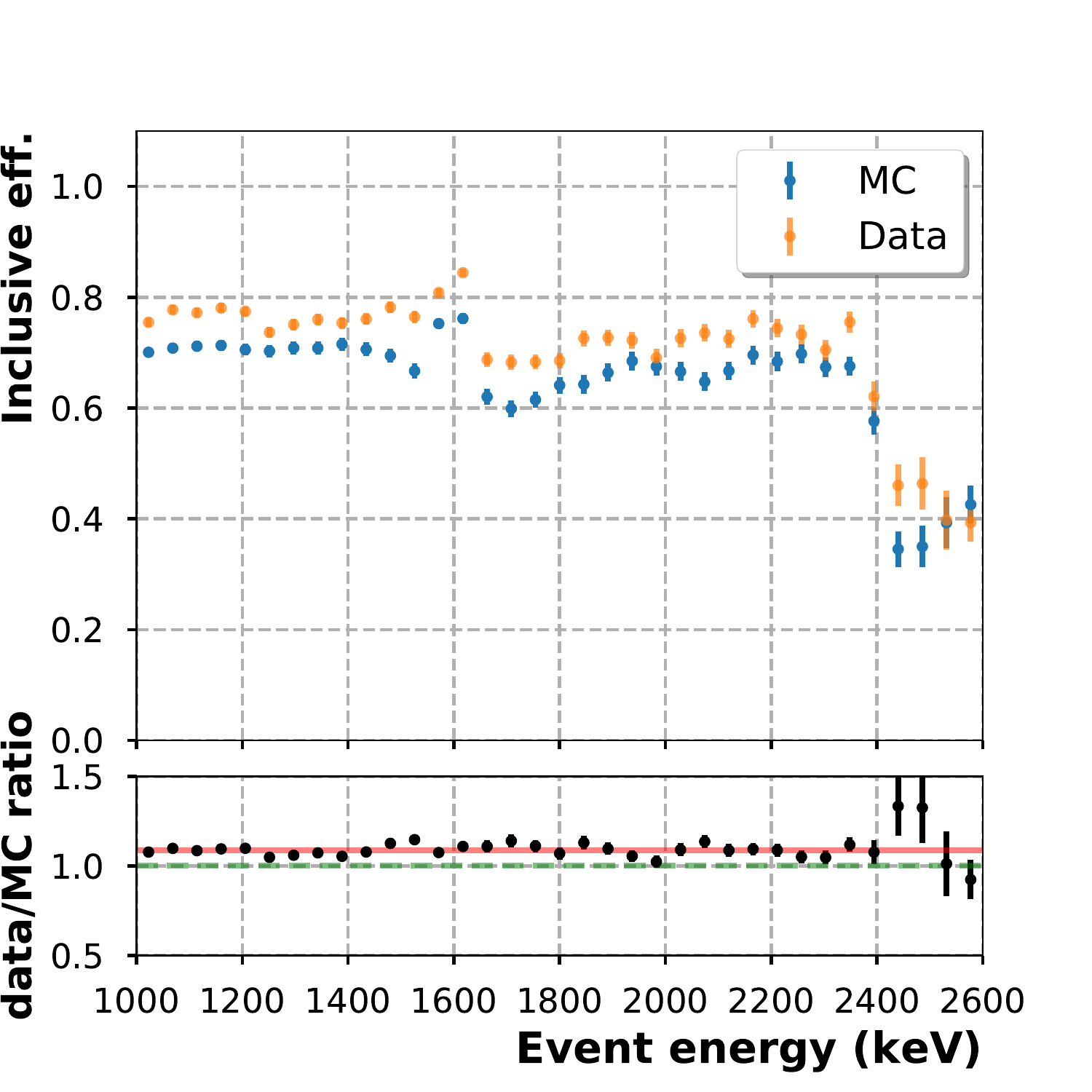}
    \includegraphics[scale=0.4]{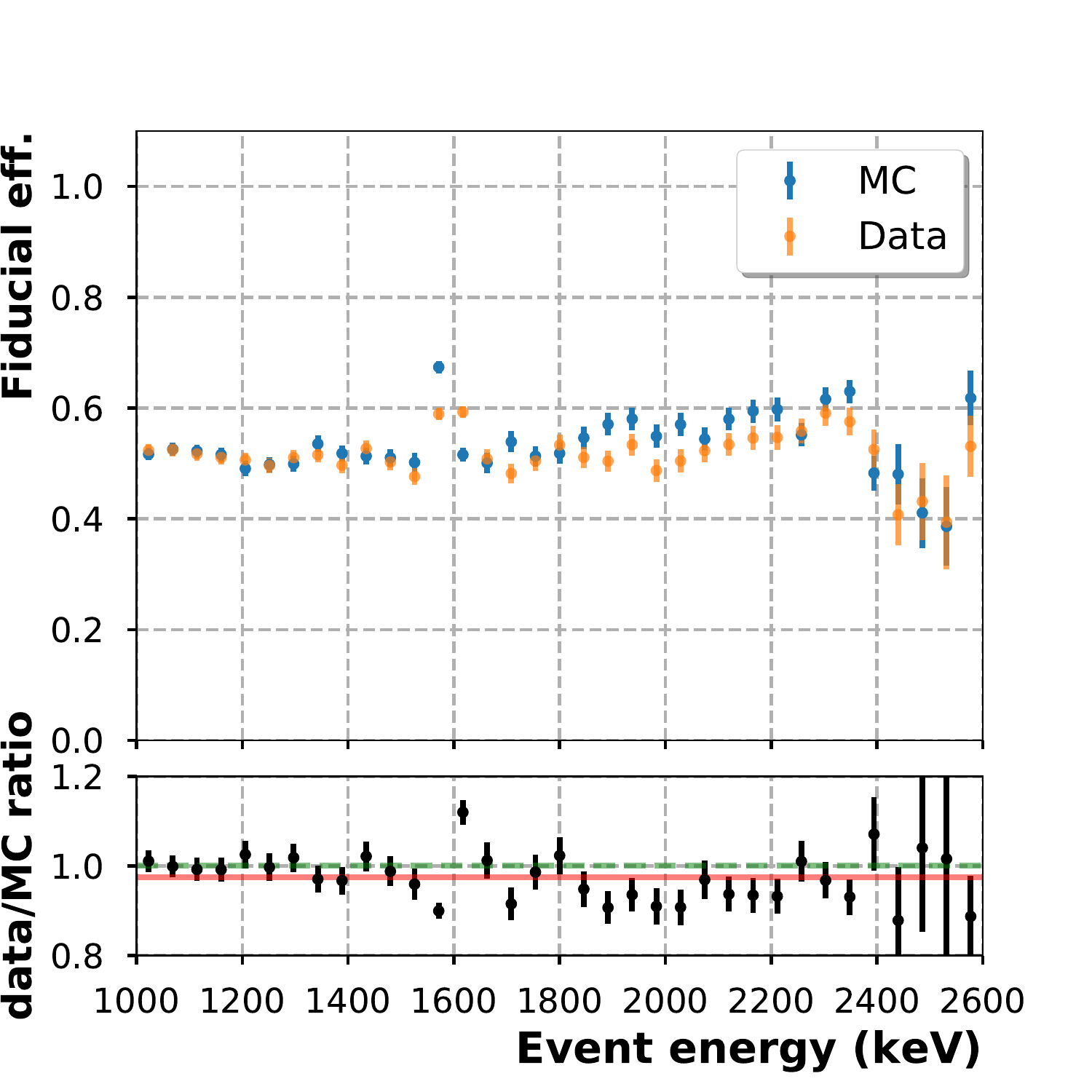}
    \caption{ Efficiency of inclusive (left) and fiducial (right) selections, as a function of the event energy. Results for data and MC are displayed with orange and blue dots, respectively. The lower panels show the ratio between data and MC fitted to a horizontal line.}
    \label{fig:inclusive_eff}
  \end{center}
\end{figure}

The efficiency of the inclusive and fiducial selections is evaluated using a calibration run where a \Th{232} source was deployed at the top of the pressure vessel. This run provided a sample of 42,788 \Tl{208} candidate events with a reconstructed energy ($E$) greater than 1000 keV, prior to the inclusive requirement. The energy reconstruction procedure is defined in Sec.~\ref{sec:backgrounddata}. Calibration data are compared to a MC simulation of 40,523 \Tl{208} decays generated at the same location, and obtained with the same reconstruction/selection procedure as data. The efficiencies of the inclusive and fiducial selections are displayed as a function of event energy in Fig.~\ref{fig:inclusive_eff}. The energy dependence of the efficiencies is well reproduced by the MC. The increase of the efficiency around 1600~keV corresponds to \Tl{208} 2615~keV gamma pair-production events, where the two 511~keV gammas produced by positron annihilation escape the detector (\Tl{208} double-escape events, in the following). For the same total energy, double-electron tracks are shorter than single-electron tracks. Hence, the corresponding probability of being properly reconstructed as a single S2 is larger, as is the probability of being fully contained in the fiducial volume. The efficiency drop above $\sim$2300~keV is due to multi-Compton events produced by the 2615~keV gammas. The relative inclusive and fiducial selection efficiencies, for $E>1000$~keV, are (74.8$\pm$0.2)\% and (52.5$\pm$0.3)\% in data, respectively, while the MC yields (68.5$\pm$0.2)\% and (53.7$\pm$0.3)\%. This reflects some level of disagreement between data and MC at the signal reconstruction stage, which is accounted for when comparing background samples in data and MC. In particular, we re-scale the background expectations according to the best-fit calibration data/MC efficiency ratios in Fig.~\ref{fig:inclusive_eff}, and propagate the uncertainty in the ratios when quoting low-background MC expectations after fiducial cuts.

\section{Fiducial background measurement}
\label{sec:backgrounddata}

Extended NEXT-White low-background operations at 10.1~bar gas pressure and with xenon depleted in \Xe{136} ($\simeq$3\% isotopic abundance) started on August 2018 and lasted until Jan 2019. All data samples were collected with the outer lead castle being closed, and with the same electric fields for the drift volume and the EL gap: $\Delta V_{drift}$ = 22.1~kV and $\Delta V_{EL}$ = 7.9~kV. The electron lifetime improved with time although fluctuations correlated with changes in the LSC HALL A temperature were observed. According to the detector operating conditions, the low-background data phase of Run-IV has been divided into three periods: Run-IVa, Run-IVb and Run-IVc. Run-IVa corresponds to 41.5 days of data taken before the radon abatement system started flushing air inside the outer lead castle. Run-IVb corresponds to data taken with radon-depleted air in the lead castle. The effective exposure of this period, which started once the RAS began stable operation, is 27.2 days. Finally, Run-IVc consists of data taken with radon-suppressed air and with the inner lead castle surrounding the pressure vessel. Data corresponding to an effective exposure of 37.9 days have been collected. Table~\ref{tab:runivdata} summarizes the data taking statistics during Run-IV. 

\begin{table}[!htb]
\caption{\label{tab:runivdata} Run-IV low-background data samples.}
\begin{center}
\begin{tabular}{ccccc}
\hline
Run period & Start Date  & Run time (day) & Triggers  & Operation conditions\\ \hline
Run-IVa    & 07-08-2018  & 41.5           &  452,407   & RAS Off, No ILC \\ 
Run-IVb    & 14-10-2018  & 27.2           &  222,498   & RAS On, No ILC\\
Run-IVc    & 29-11-2018  & 37.9           &  302,084  & RAS On, ILC \\ \hline
\end{tabular}
\end{center}
\end{table}

\begin{figure}
  \begin{center}
    \includegraphics[scale=0.80]{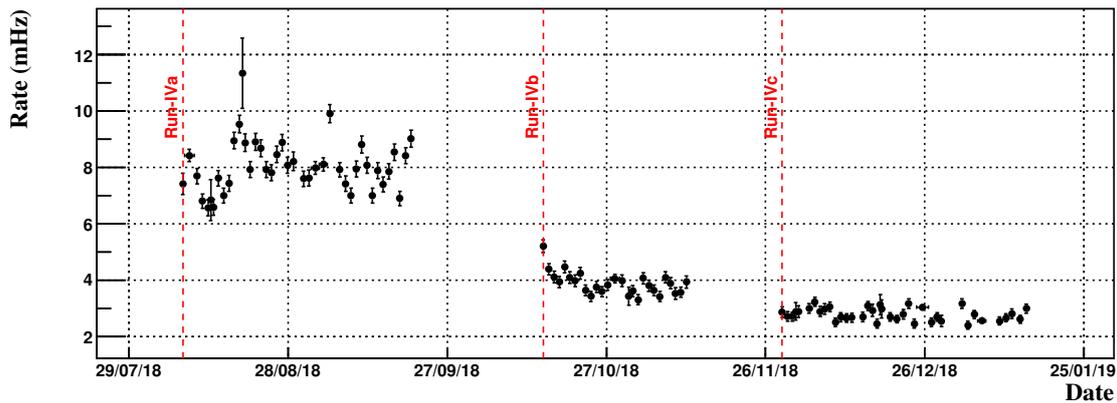}
    \caption{Fiducial background rate as a function of data taking calendar day. Vertical dashed lines mark the start time of Run-IVa, Run-IVb and Run-IVc.}
    \label{fig:rate}
  \end{center}
\end{figure}

In order to evaluate the total background in NEXT-White before any $\beta\beta$ selection cuts, both the rate and the energy spectrum have been measured in the three Run-IV periods. The fiducial background rates are listed in Tab.~\ref{tab:rate}, once an energy threshold of 600~keV has been applied. The fiducial background rate as a function of time is also shown in Fig.~\ref{fig:rate}. The DAQ system dead-time has been computed on a daily basis, as it is correlated to the \Kr{83m} rate. The amount of \Kr{83m} decays in the active volume evolves with time according to the half-life of the parent \Rb{83} source and the flux of the gas system, leading to variations of the DAQ dead-time within 2\%: the higher the \Kr{83m} activity inside the detector, the higher the DAQ dead-time. The integrated DAQ live-time for the entire Run-IV is found to be (94.80$\pm$0.04)\%. The trigger efficiency for events above 600~keV has also been measured to be (77.8$\pm$0.9)\%. The significant trigger inefficiency is due to the coincidence time window between the two PMTs used to trigger the DAQ system. The rates presented in Tab.~\ref{tab:rate} and Fig.~\ref{fig:rate} are corrected for the DAQ dead-time and the Run-IV trigger inefficiency, assigning a systematic uncertainty of 0.9\%. The configuration of the trigger has been improved in Run-V and the trigger efficiency is now close to 100\%.

\begin{table}[!htb]
\caption{\label{tab:rate} Run-IV background rates for events with energy above 600 keV.}
\begin{center}
\begin{tabular}{ccc}
\hline
Run period & Inclusive rate (mHz) & Fiducial rate (mHz) \\ \hline
Run-IVa    &  19.09$\pm$0.08$_{\rm stat}\pm$0.17$_{\rm syst}$  &  8.00$\pm$0.05$_{\rm stat}\pm$0.07$_{\rm syst}$ \\ 
Run-IVb    &  11.28$\pm$0.08$_{\rm stat}\pm$0.10$_{\rm syst}$  &  3.90$\pm$0.05$_{\rm stat}\pm$0.04$_{\rm syst}$ \\
Run-IVc    &   8.97$\pm$0.06$_{\rm stat}\pm$0.08$_{\rm syst}$  &  2.78$\pm$0.03$_{\rm stat}\pm$0.03$_{\rm syst}$ \\ \hline
\end{tabular}
\end{center}
\end{table}

\begin{figure}
  \begin{center}
    \includegraphics[scale=0.33]{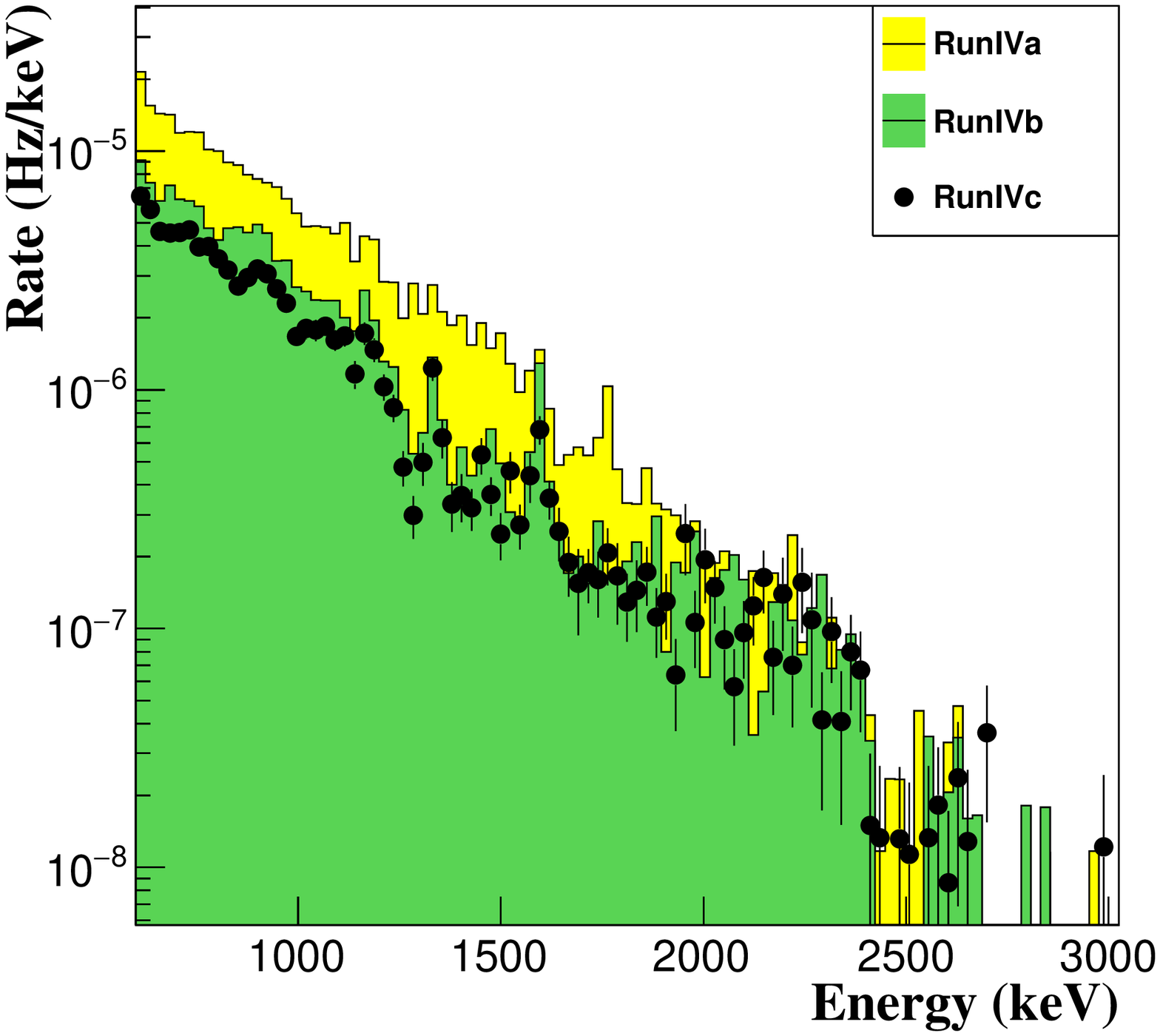}
    \includegraphics[scale=0.33]{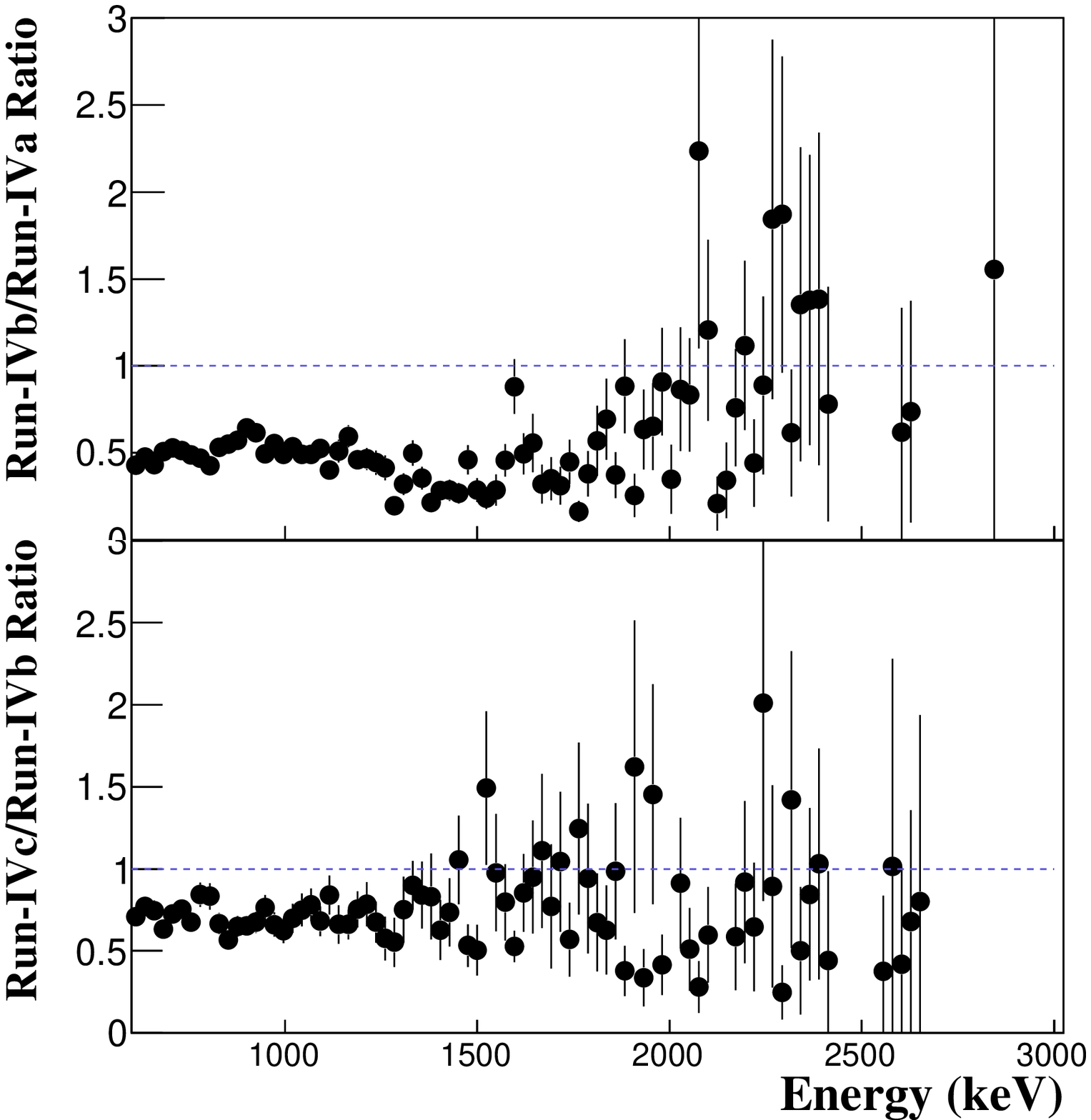}
    \caption{Fully corrected energy spectra of the fiducial background samples collected during Run-IV. Left: spectra from Run-IVa, Run-IVb and Run-IVc superimposed. For clarity, the statistical error bars in Run-IVa, Run-IVb are not shown. Right: ratio between Run-IVb and Run-IVa (top) and between Run-IVc and  Run-IVb (bottom).}
    \label{fig:energy}
  \end{center}
\end{figure}

The energy of the selected events has been reconstructed as follows. First, the PMT charge associated to each reconstructed 3D hit in the event is separately corrected for electron attachment. The electron lifetime assumed for the correction is derived from the $^{83m}$Kr data collected within a $\sim$24-hour period, where time variations are also taken into account. The second step consists of a geometrical XY correction of the detector response depending on the hit XY position. The correction relies on a XY energy map obtained also from the $^{83m}$Kr data within the same 24-hour period. A preliminary linear energy scale is applied to convert the sum of the hit corrected charges (in photo-electrons) into event energy (in keV). The conversion factor is estimated from the 41.5~keV electron-conversion $^{83m}$Kr peak, accounting also for sub-percent time variations in the light yield during each 24-hour period. The final energy scale is obtained from high-energy calibration runs, deploying \Cs{137} and \Th{232} sources, taken before (after) the start (end) of Run-IV. The \Cs{137} photo-peak (662~keV) and the \Tl{208} double-escape peak (1592 keV) and photo-peak (2615~keV) are used to define a linear scale yielding residuals below 0.4\%. Figure~\ref{fig:energy} shows the energy spectra of the fiducial background samples in Run-IV, for an energy above 600~keV. Despite the limited exposure, the characteristic lines of $^{208}$Tl (1592~keV), \Bi{214} (1764 and 2204~keV), \Co{60} (1173 and 1333~keV) and \K{40} (1461~keV) isotopes are visible.

\begin{figure}
  \begin{center}
    \includegraphics[scale=0.66]{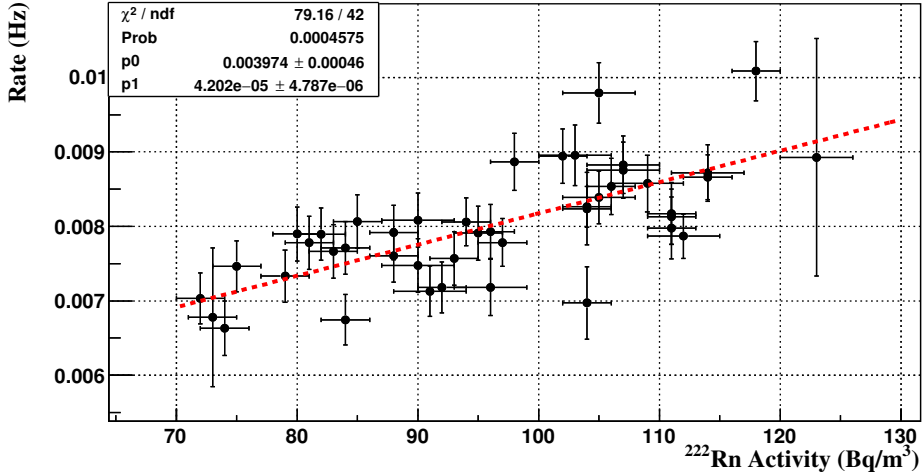}
    \caption{Run-IVa fiducial background rate versus airborne radon activity. A linear fit extrapolation to zero-Rn-activity yields an expected background rate of 3.97$\pm$0.46 mHz.}
    \label{fig:rncorr}
  \end{center}
\end{figure}

The background rate in Run-IVa has decreased by a factor of 1.7 with respect to the earlier pilot background run taken in 2017 (Run-II), despite the pressure increase from 7.2 to 10.1~bar. This background rate reduction confirms the expected improvement in detector radiopurity introduced by the replacement of the resistor chain of the field cage and the PMT bases. However, the rate variations in time (not consistent with statistical fluctuations) observed for Run-IVa in Fig.~\ref{fig:rate} are a clear indication of a time-dependent background source, thereby not related to radio-impurities of the detector materials. From the analysis of the correlation of the background rate with the level of airborne radon at the LSC, it is concluded that such variations are due to a significant contribution of $^{222}$Rn decays within the volume of the lead castle. Using the radon activity data provided by an AlphaGUARD detector (Bertin Instruments), the correlation is quantified in Fig.~\ref{fig:rncorr} by means of a linear fit. From this fit, an expectation of the fiducial background rate in NEXT-White for a zero Rn activity is derived: (3.97$\pm$0.46)~mHz.    

The effect of the RAS in Run-IVb is clearly visible in Fig.~\ref{fig:rate} and Fig.~\ref{fig:energy}. After a period of a few days where the background rate decreases as the the remaining $^{222}$Rn inside the outer lead castle decays, the fiducial background rate becomes stable and a reduction of a factor of 2.1 with respect to Run-IVa is reached. The comparison of the energy spectra in Run-IVa and Run-IVb around the 1764~keV gamma line of \Bi{214}, a progeny of \Rn{222}, positively identifies this reduction as due to \Rn{222} suppression. In addition, the amplitude of the \Tl{208} double-escape line at 1592~keV is not reduced. The consistency between the background rate measurement in Run-IVb, (3.90$\pm$0.06)~mHz, and the zero Rn activity background extrapolation from Run-IVa, (3.97$\pm$0.46)~mHz, implies that the RAS allows for operation of the NEXT-White detector in an environment virtually free of airborne Rn. In particular, this validates the assumption of a negligible external radon-induced background in the evaluation of the physics case of the NEXT-100 detector~\cite{Martin-Albo:2015rhw}.  

The main goal of the ILC installed between Run-IVb and Run-IVc is to provide further shielding against background contributions coming from the outer lead castle volume. A reduction of about an order of magnitude in these contributions is expected according to Monte-Carlo simulations. The data taken during Run-IVc offer a handle to understand the overall background budget, by means of the comparison with Run-IVa and Run-IVb periods. As shown in Fig.~\ref{fig:rate} and Fig.~\ref{fig:energy}, the Run-IVc data shows a reduction in the fiducial background rate of about 40\% with respect to Run-IVb. This implies that Run-IVa and Run-IVb suffer from a significant contribution of external backgrounds not related to airborne radon ($\sim$1 mHz). Although the origin of this external background is unclear, the main candidates are the castle structure paint and the rails and mechanical structures inside the outer lead castle. According to Fig.~\ref{fig:energy}, the contributions from \Tl{208} and \Bi{214} are clearly reduced. On the other hand, the amplitude of the \Co{60} lines remain essentially the same, pointing to an internal origin of this source of background. Beyond the reduction of the overall background rate, it is worth remarking upon the stability of the rate over time. As Run-IVc data were taken with the same operating conditions as for the enriched \Xe{136} run (Run-V), the observed background is used in Sec.~\ref{sec:backgroundfit} to validate the NEXT background model, and in Sec.~\ref{sec:betabetasearch} to estimate the backgrounds in the \bbtwonu and \bbnonu analyses.

\section{Radiogenic background model}
\label{sec:backgroundmodel}

The expected background budget in NEXT-White is derived from a detailed background model accounting for different isotopes and detector volumes. The model relies on the extensive radiopurity measurements campaign conducted by the NEXT collaboration \cite{Alvarez:2012as,Alvarez:2014kvs,Cebrian:2017jzb}. A total of 44 detector materials have been considered, screening their \Bi{214}, \Tl{208}, \K{40} and \Co{60} contributions. The measurement technique employed for most materials is gamma spectroscopy with high-purity Germanium detectors of the LSC Radiopurity Service. In order to reach sensitivities below 1~mBq/kg, mass spectroscopy techniques (ICPMS, GDMS) have also been used for some detector materials, namely copper, lead, steel, and high-density polyethylene. The background model conservatively assumes the 95\% CL upper limits obtained for each (isotope, material) combination where the specific activity could not be quantified, while the measured activity central values are used otherwise. These specific activity assumptions are then multiplied by the material quantities to obtain the total background activity assumptions, in mBq. The material quantities are obtained from the as-built engineering drawings of NEXT-White and the known material densities. A contribution from Rn-induced \Bi{214} on the cathode surface is also considered, according to the measurements performed in \cite{Novella:2018ewv}. In addition, the contribution from the \bbtwonu of \Xe{136}, whose fraction in the depleted Xe used in Run-IV is (2.6$\pm$0.2)\%, is also incorporated into the model. 

According to these radiopurity measurements, a full GEANT4-based Monte-Carlo simulation has been performed. The screened materials are associated to 22 GEANT4 volumes describing the components of the NEXT-White detector, with one or more materials assigned to each volume. A detection efficiency is estimated for each (isotope, GEANT4 volume) combination. The detection efficiency is defined as the number of radioactive decays depositing at least 400~keV in the detector active volume, divided by the total number of radioactive decays. The highest detection efficiencies are obtained for the innermost volumes, such as the cathode grid, the Teflon light tube and the anode quartz plate. On the other hand, the outermost simulated volumes, particularly the lead-based shielding structure, have detection efficiencies as low as $10^{-7}$. The simulated shielding geometry includes the ILC and its steel structure, so the derived model can be compared with Run-IVc data. The gas pressure assumed in simulations is 10.1~bar, also comparable with Run-IVc data. Overall, the model contains 84 background sources, one for each (isotope, GEANT4 volume) contribution considered. 

\begin{table}[!htb]
  \caption{\label{tab:backgroundmodel}Most important background contributions in NEXT-White according to our model, for events depositing more than 400~keV of energy in the TPC active volume. The background isotope, GEANT4 volume, fit volume, total activity, detection efficiency and expected event rate are listed for each background contribution, with contributions ordered by decreasing event rate.}
\begin{center}
\begin{tabular}{cccrc|r}
  \hline
  Isotope  & G4 Volume & Fit Volume & Activity             & Efficiency & Rate \\
           &             &            & (mBq)                &            & (mHz) \\ \hline
  \Bi{214} & Cathode      & Cathode    & $3.10\times 10^{0}$  & $6.50\times 10^{-1}$  & $2.02$ \\
  \Bi{214} & Drift Tube   & Other      & $<3.10\times 10^{1}$  & $3.14\times 10^{-2}$ & $<0.97$ \\
  \Co{60}  & Vessel       & Other      & $2.00\times 10^{3}$  & $4.16\times 10^{-4}$ &  $0.83$ \\
  \K{40}   & Dice Board   & Anode      & $4.07\times 10^{2}$  & $1.82\times 10^{-3}$ &  $0.74$ \\
  \K{40}   & Drift Tube   & Other      & $<1.16\times 10^{2}$  & $5.68\times 10^{-3}$ & $<0.66$ \\
  \Tl{208} & Drift Tube   & Other      & $<1.17\times 10^{1}$  & $4.99\times 10^{-2}$ & $<0.58$ \\
  \Co{60}  & Drift Tube   & Other      & $<8.30\times 10^{0}$  & $4.06\times 10^{-2}$ & $<0.34$ \\
  \Tl{208} & Extra Vessel & Anode      & $3.69\times 10^{3}$  & $8.67\times 10^{-5}$ &  $0.32$ \\
  \Co{60}  & PMT Body     & Cathode    & $4.56\times 10^{1}$  & $6.49\times 10^{-3}$ & $0.30$ \\ \hline
  \multicolumn{5}{c|}{Other} & $<1.92$ \\ \hline
  \multicolumn{5}{c|}{Total} & $<8.68$ \\ \hline
\end{tabular}
\end{center}
\end{table}

Table~\ref{tab:backgroundmodel} shows the most important background contributions for $E>400$~keV in NEXT-White according to our model. Background sources contributing $>3\%$ of the total event rate above 400~keV are listed. For each background source given by a specific (isotope, GEANT4 volume) combination, the total background activity, detection efficiency and event rate is given. The total activity is indicated with a ``less-than'' sign if it is based on a material screening upper limit. The event rate is the product of total activity times detection efficiency. The most important background contribution above 400~keV is expected to be \Bi{214} decays from the cathode grid, induced by internal radon contamination \cite{Novella:2018ewv}. Contaminants in the Teflon light tube (``drift tube''), in the pressure vessel and in the kapton printed circuit boards used for the tracking plane (``Dice Board'') are also important.

\begin{figure}
  \begin{center}
    \includegraphics[width=0.80\textwidth]{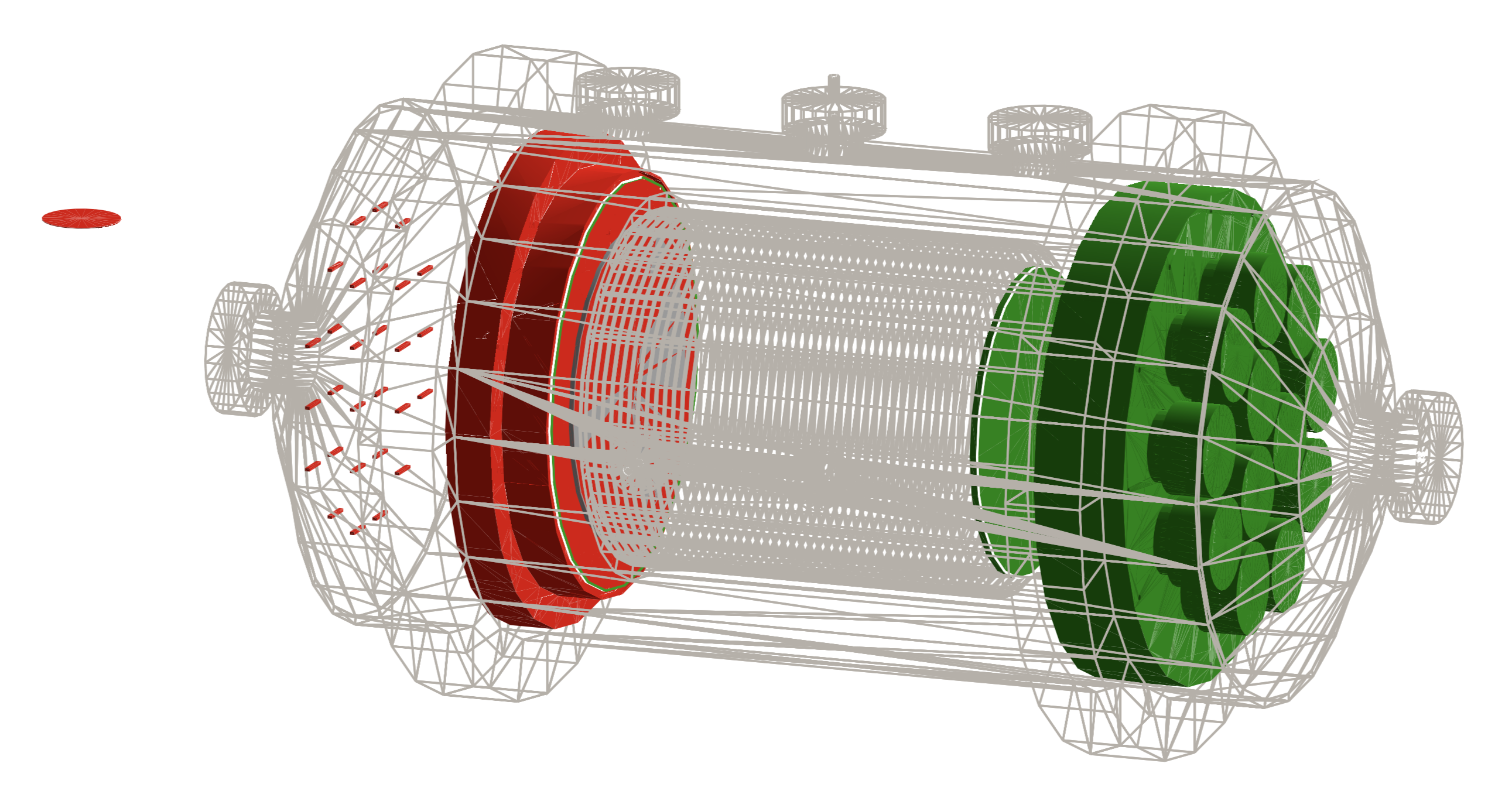} 
    \caption{GEANT4 description of the NEXT-White geometry. The red, green and grey volumes in the figure are included into the ``Anode'', ``Cathode'' and ``Other'' fit volume categories, respectively.}
    \label{fig:nextwhite_fitvolumes}
  \end{center}
\end{figure}

Table~\ref{tab:backgroundmodel} also indicates the associated fit volume for each background source. As we will see in Sec.~\ref{sec:backgroundfit}, the low-background data provide some handle to identify the spatial origin of the events, but not enough to separately constrain 22 GEANT4 volumes. Instead, for background fitting purposes, the background sources are grouped into three spatial categories: ``Anode'', ``Cathode'' and ``Other''. The ``Anode'' and ``Cathode'' categories include all GEANT4 volumes placed in, or near to, the two detector end-caps. The ``Other'' category include inner volumes in the detector barrel region, the pressure vessel and external backgrounds such as the ones emanating from the shielding structure. For a visual representation of the three fit volume categories, see Fig.~\ref{fig:nextwhite_fitvolumes}.

\begin{figure}
  \begin{center}
    \includegraphics[width=0.49\textwidth]{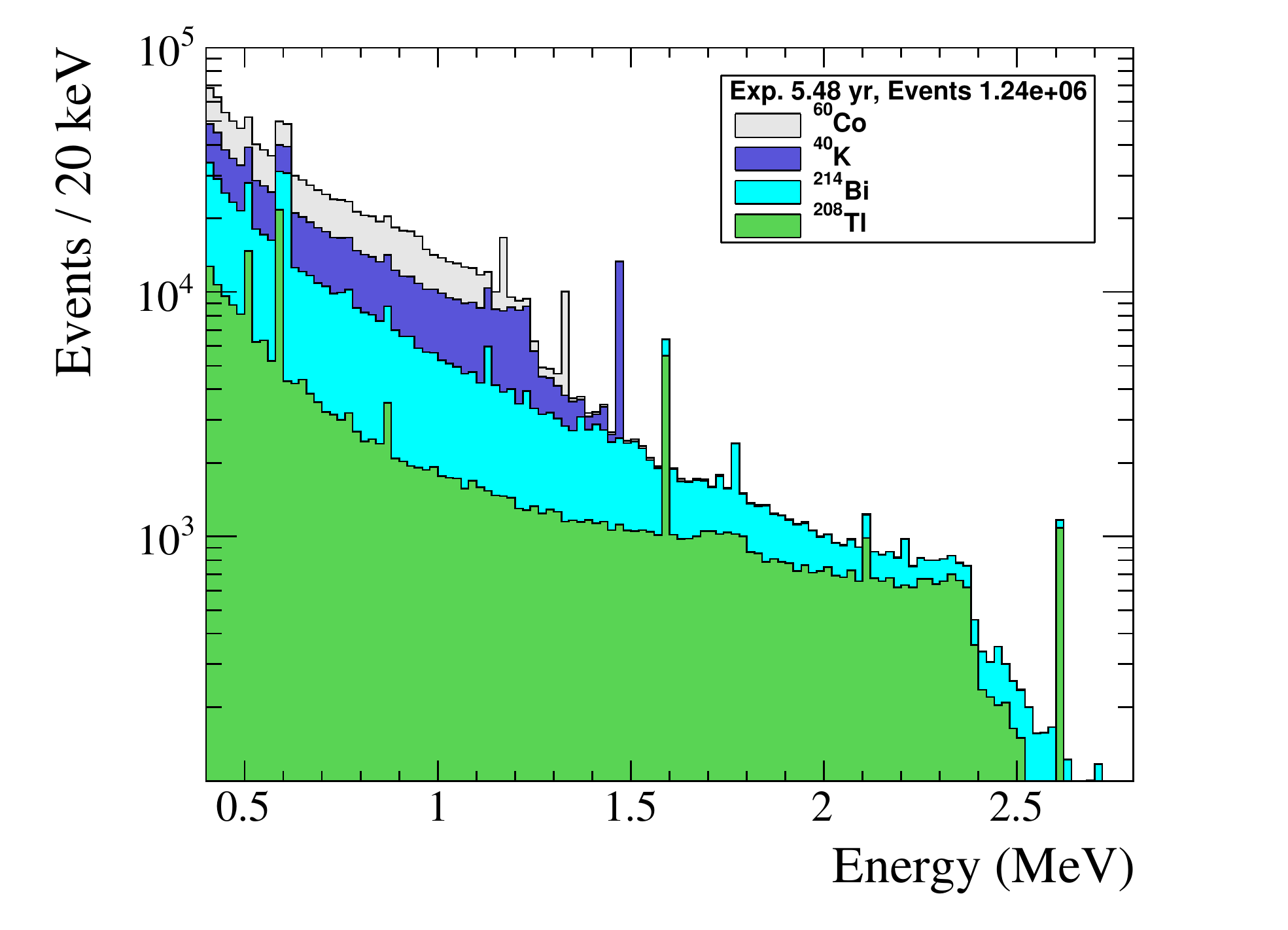} \hfill
    \includegraphics[width=0.49\textwidth]{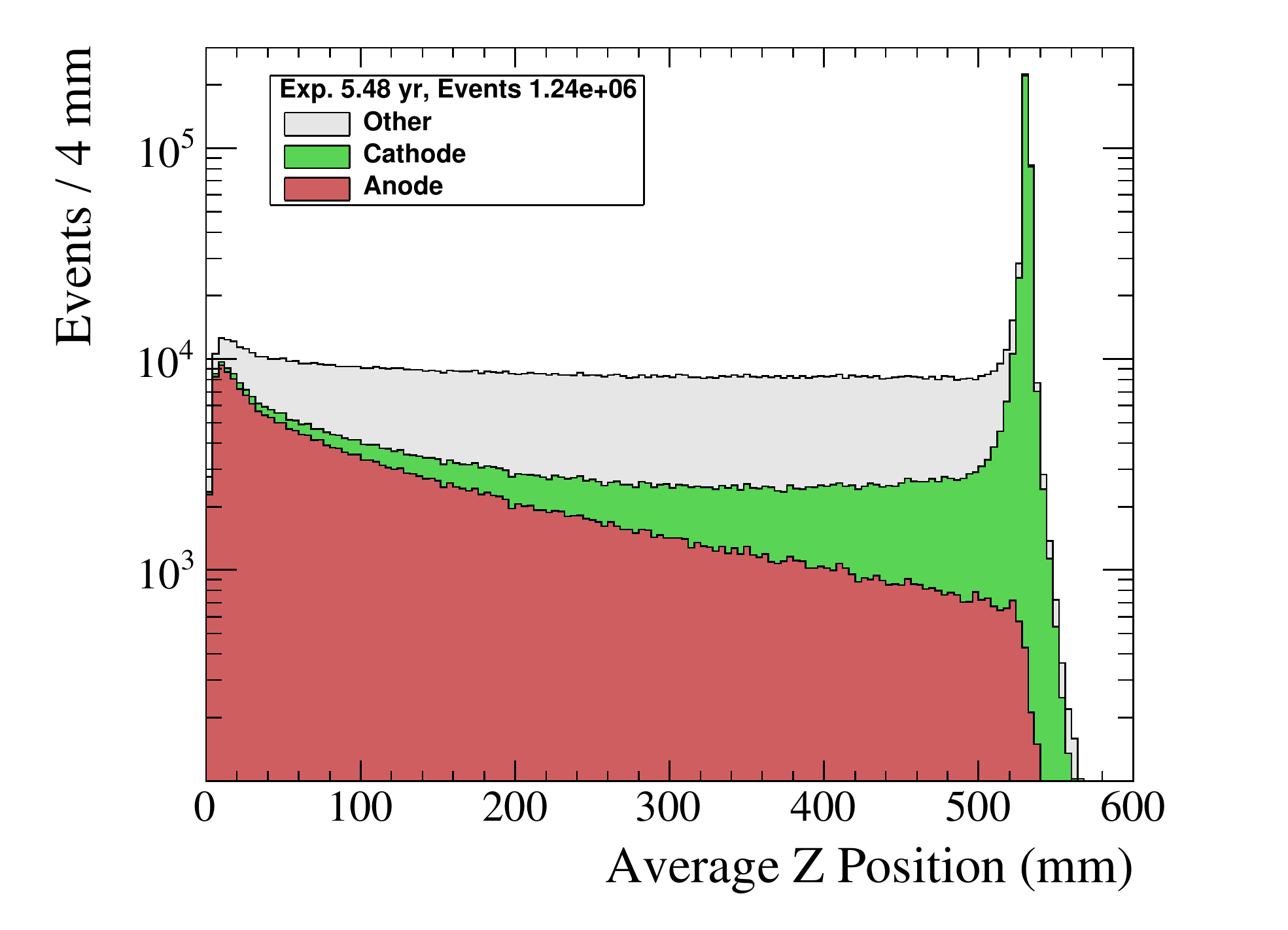}
    \caption{Background model expectations for all events depositing $>400$~keV in the TPC active volume. The stacked histogram in the left panel shows the event energy distribution, broken by isotope type. The stacked histogram in the right panel shows the event average z position, separated by fit volume type.}
    \label{fig:bgr_model}
  \end{center}
\end{figure}

Figure~\ref{fig:bgr_model} shows the background model expectations for all events depositing $>400$~keV in the TPC active volume, prior to event reconstruction and selection. The low-energy portion of the event energy distribution is dominated by \Co{60} and \K{40} activities, extending up to 1333~keV and 1461~keV, respectively. The \Bi{214} and \Tl{208} contributions account for the high-energy part. The event z (drift) position is also shown in Fig.~\ref{fig:bgr_model}, where we show the charge-weighted average over all GEANT4 TPC active volume hits in the event. As expected, the ``Anode'' and ``Cathode'' fit volume contributions are peaked at low ($z\simeq 0$) and high ($z\simeq 530$~mm) drift positions, respectively, while the ``Other''  component is more uniform.

Overall, about $10^{11}$ background events have been generated with our GEANT4-based simulation. From those, a sample of 1.5 million background events with visible energy $>400$~keV in the TPC active volume is obtained and processed through the entire simulation/reconstruction chain, corresponding to an exposure of 5.48~years. The GEANT4 events are then processed to mimic the electronic effects (shaping of the electronics, noise, digitization), so that the corresponding raw waveforms can be compared to the ones collected by the DAQ system. Then, the Monte-Carlo events are passed through the same reconstruction and corrections steps as described for real data, and through the same fiducial selection, see Secs.~\ref{sec:fiducialsel} and \ref{sec:backgrounddata}. The expected background rate after full reconstruction and fiducial selection in Run-IVc is (0.489$\pm$0.002$_{\rm stat}\pm$0.004$_{\rm syst}$)~mHz for $E>1000$~keV, where the systematic error is due to the non-perfect knowledge of the inclusive and fiducial selections, see Sec.~\ref{sec:fiducialsel}.

\section{Background characterization}
\label{sec:backgroundfit}

A detailed comparison of Run-IVc background data and the Monte-Carlo background model has been performed. Beyond the validation of the model, such a comparison helps to identify the main contributions to the total background budget by exploiting the energy and spatial information of the events. In turn, the results allow for the tuning of the background expectations prior to $^{136}$Xe double beta decay searches in NEXT, as done in Sec.~\ref{sec:betabetasearch}.

\begin{figure}
  \begin{center}
    \includegraphics[scale=0.37]{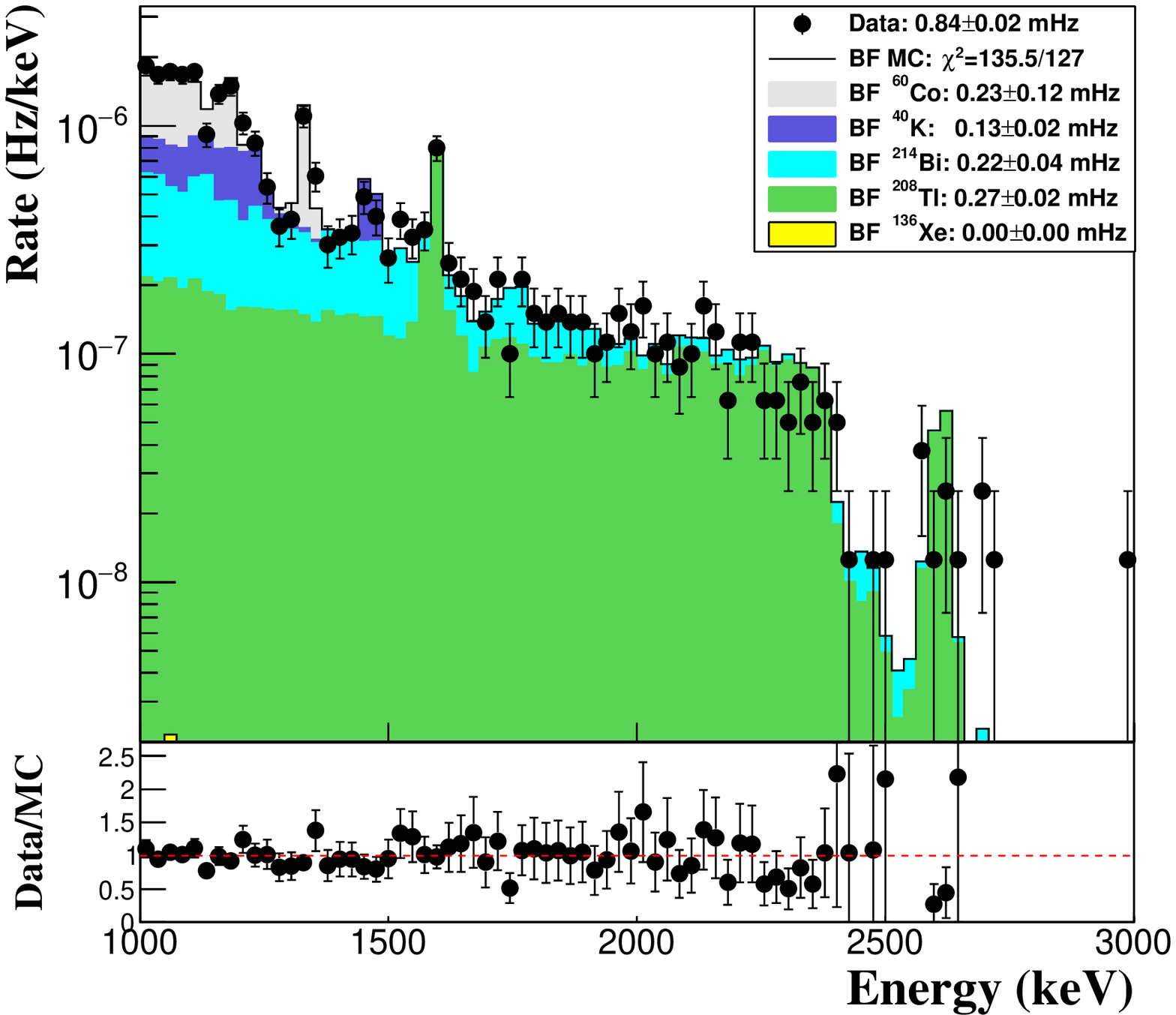}
    \includegraphics[scale=0.37]{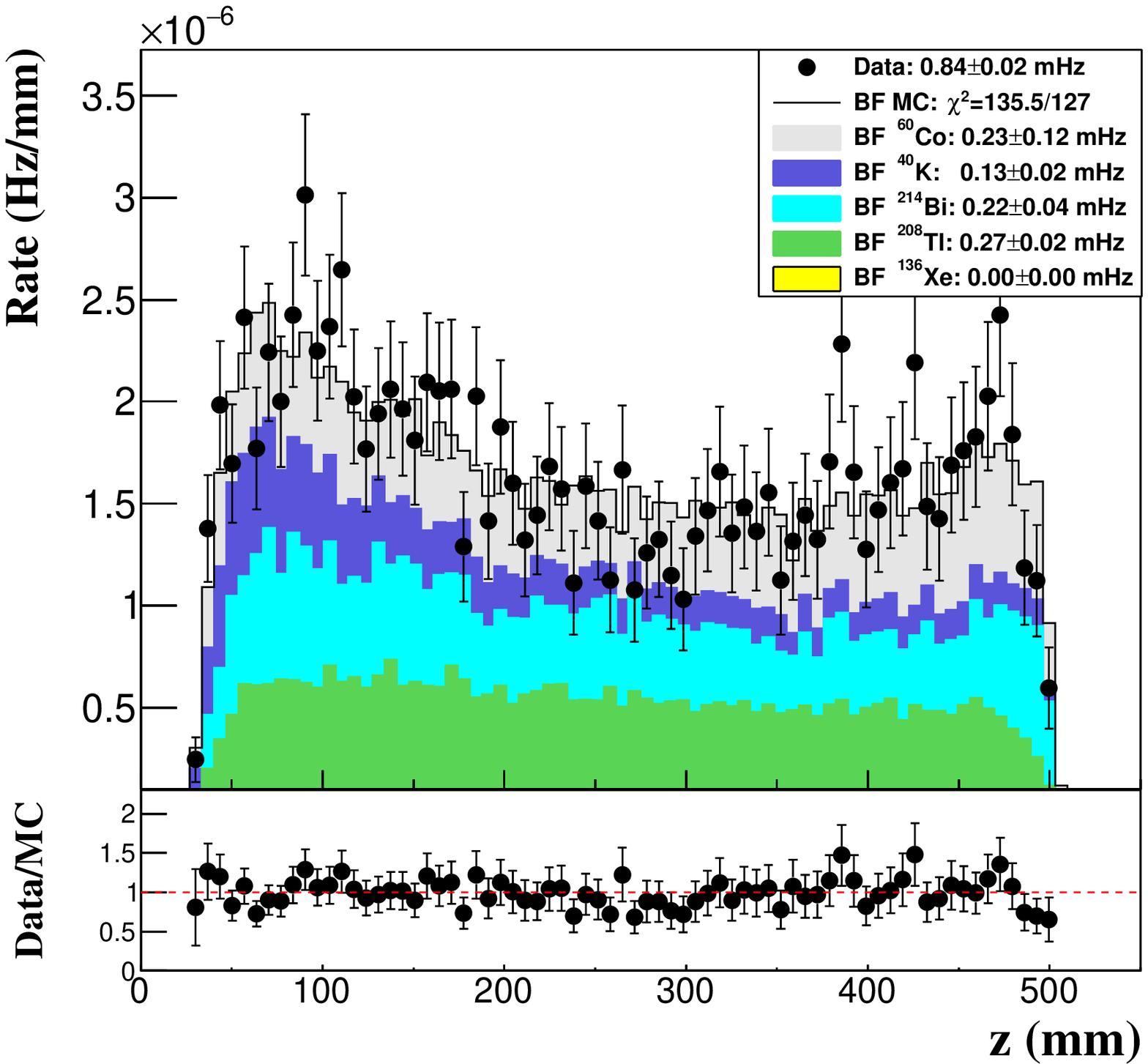}
    \caption{ Run-IVc background fit. Data (black dots) are superimposed to the best-fit background model expectation (solid histograms), for which the different isotopes contributions are shown. The displayed isotope rates are obtained by propagating the scale factor best-fit values of the three effective volumes for each isotope.}
    \label{fig:bgfit}
  \end{center}
\end{figure}

In order to normalize the different contributions to the background model so that it matches the data, an effective fit has been performed in the 1000--3000 keV range. The fit consists of the minimization of a maximum extended likelihood, considering both energy and z (drift) coordinate information. Three effective background volumes are considered in the model, as discussed in Sec~\ref{sec:backgroundmodel}. The rationale for the definition of these effective volumes is to exploit the z-dependence observed in the background data. The fit considers the contribution of four isotopes (\Bi{214}, \Tl{208}, \K{40} and \Co{60}) from the 3 effective volumes, resulting in a total of 12 fit parameters that provide normalization factors with respect to the nominal model predictions. Since the contribution of \Xe{136} is negligible, its normalization has been fixed to the nominal value. The comparison of the fiducial background in Run-IVc and the best-fit model is shown in Fig.~\ref{fig:bgfit}. With a reduced chi-square of $\chi^2$/ndof=1.07 (p-value = 0.29), the best-fit reproduces reasonably well the energy spectrum and the z distribution. Considering the contributions from the three effective volumes and their correlations, the best-fit overall normalization factors for \Co{60}, \K{40}, \Bi{214} and \Tl{208} are 2.70$\pm$0.22, 0.76$\pm$0.11, 2.21$\pm$0.37 and 1.95$\pm$0.15, respectively. The corresponding best-fit rates for each one of the isotopes are displayed in the legend of Fig.~\ref{fig:bgfit}. The precision on these rates range from 8\% (\Tl{208} and \Co{60}) to 17\% (\Bi{214}). Summing over all isotopes, the overall scale factor of the expected total rate is 1.72$\pm$0.04 with respect to the nominal background prediction.

\begin{figure}
  \begin{center}
    \includegraphics[scale=0.37]{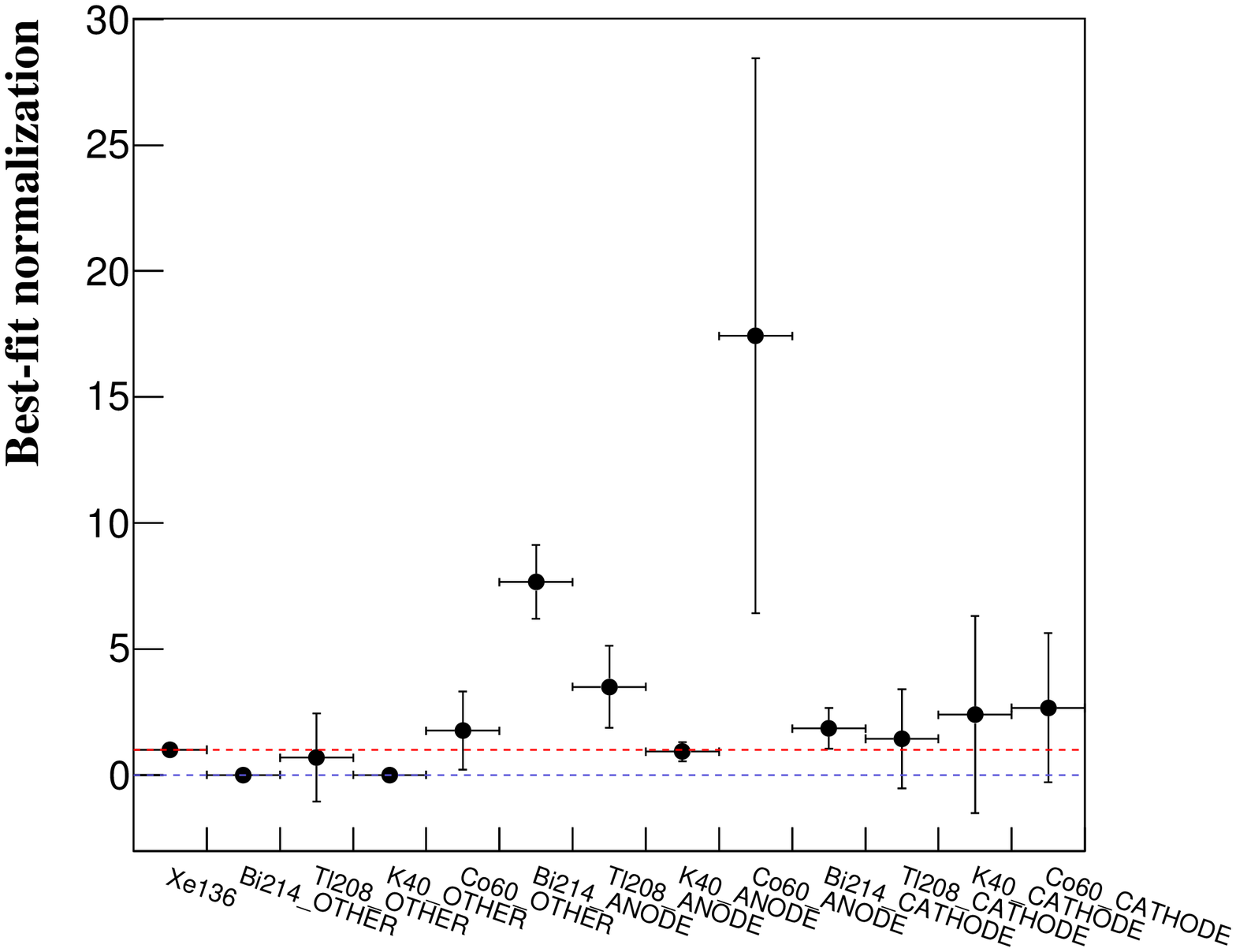}
    \includegraphics[scale=0.37]{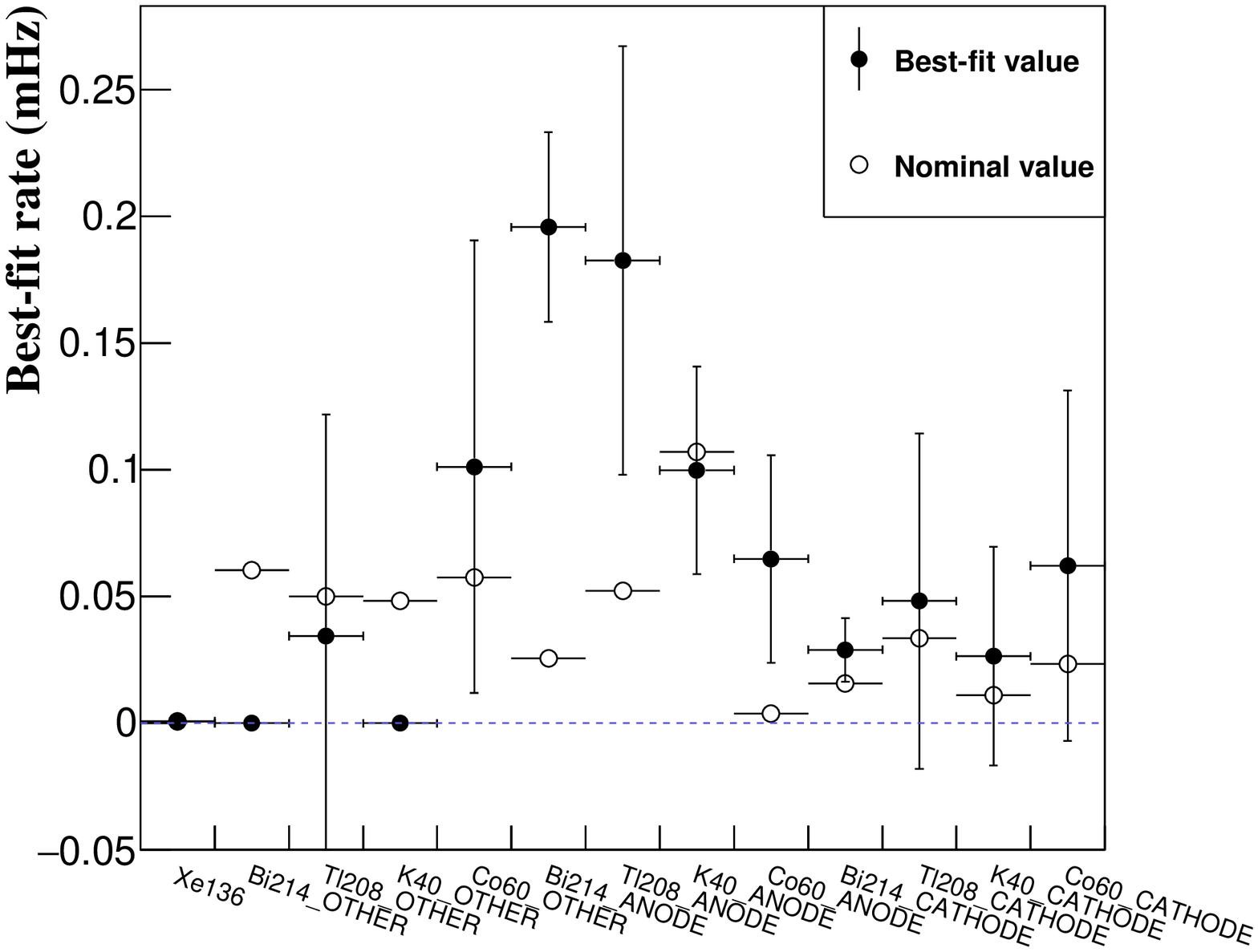}
    \caption{ Best-fit values of the 12 parameters considered in the Run-IVc background fit. Left: scaling factors of the 4 isotopes in the three effective volumes with respect to the nominal values in the background model. Right: corresponding nominal (empty circles) and best-fit (solid circles) rates. }
    \label{fig:bgbfvals}
  \end{center}
\end{figure}

Central values and errors for the 12 fit parameters are shown in Fig.~\ref{fig:bgbfvals}, in terms of the normalization factors and the corresponding rates. These values provide relevant information about the origin of the different sources of background. The excess of events in the low energy and low-z regions is compensated mostly by contributions from \Co{60}, \Bi{214} and \Tl{208} from the anode, yielding normalization factors of 17.4$\pm$11.0, 7.7$\pm$1.5 and 3.5$\pm$1.6, respectively. As a consequence, the anode region becomes the dominant contributor to the total background budget. These large deviations from the background model point to a possible unaccounted background source in the anode region which is currently under investigation. In addition, it must be noticed that the fit is not sensitive to all fit parameters. In particular, the \Bi{214} and \K{40} contributions from the ``Other" volume converge to the physical limit of 0~mHz. There are two possible reasons for this. First, the \Bi{214} and \K{40} contributions are dominated by their ``Cathode'' and ``Anode'' volume contributions, respectively, with little sensitivity to a sub-dominant ``Other'' volume contribution. Second, and according to Tab.~\ref{tab:backgroundmodel}, these are precisely the two most important nominal background contributions that are based on radio-purity screening upper limits as opposed to actual measurements. It is therefore reasonable to expect that these fit parameters converge to values below one.

Data below 1000~keV are not considered in our current background fit. The reason is twofold. On the one hand, the inclusion of 600--1000~keV events deteriorates somewhat the goodness-of-fit, from $\chi^2$/ndof=1.07 to 1.48. On the other hand, long-lived isotopes produced by cosmogenic activation are known to contribute in this energy region, beyond the four isotopes considered in our current model. With the current limited exposure, no additional isotope has been unambiguously identified thus far, see Fig.~\ref{fig:energy}. As more low-background data are collected, the background model is expected to be completed with additional isotopes and extended toward lower energies.

In summary, Run-IVc data validate the detector-induced background model of the NEXT experiment in the 1000--3000~keV energy range. The nominal normalization of the model, derived from the screening of the detector materials, reproduces the total background rate to better than a factor of two. After fitting the model to the data sample, the best-fit normalization values for the different background contributions allow to reproduce the observed total rate within 2\%, as well as the energy and spatial distributions of the events. This implies that the model can be safely used to estimate the expected backgrounds in NEXT double beta decay searches.

\section{Double-electron topological selection}
\label{sec:topology}

The \bb analyses in NEXT rely on the selection of double-electron tracks by means of their characteristic topological signature. When traveling through xenon gas, charged particles suffer from multiple scattering and lose their energy at about a constant rate, until they become non-relativistic and come to rest. At that point, the energy loss per unit path length increases, yielding a high energy deposition in a compact region (Bragg peak). Thus, a \bb decay is reconstructed as a single continuous track with energy \emph{blobs} at both track extremes. On the contrary, background events may consist either of multi-track events, or single-track events produced by single electrons. A single-electron background event consists of a track ending with only one energy blob. These significant differences between signal and background topologies are exploited to reduce the backgrounds in the \bb decay searches, as shown in \cite{Ferrario:2015kta,Ferrario:2019kwg,Renner:2016trj}.  

In order to optimize the performance of the double-electron (signal) selection, a high-level track reconstruction is applied to the hits associated to the S2 signals. First, the hits are grouped into 3D voxels of equal size. Then, tracks are defined according to the connectivity of the voxels, following a "Breadth First Search'' algorithm which also identifies the extremes and the total length \cite{Ferrario:2015kta,Ferrario:2019kwg}. Finally, blob candidates are found by integrating the energy of the hits contained in a sphere centered at the end-points of the tracks. For the reconstruction of the Run-IVc data and MC, a radius of 21~mm is considered. In the current analysis, the double-electron selection is implemented with three cuts. First, events with only one reconstructed track are selected (hereafter, single track cut). Second, the extremes of the track are required not to overlap (hereafter, blob overlap cut). This implies that the blob candidates at both end-points do not share hits with a total energy above 1~keV. This requirement is particularly relevant for low-energy events producing short tracks. Finally, a minimum energy cut is applied to the energy of both blob candidates so that they are identified as actual Bragg peaks (hereafter, blob energy cut). To enhance the efficiency of the blob energy cut, the energy threshold is defined as a function of the track energy.                

\begin{figure}
  \begin{center}
    \includegraphics[scale=0.4]{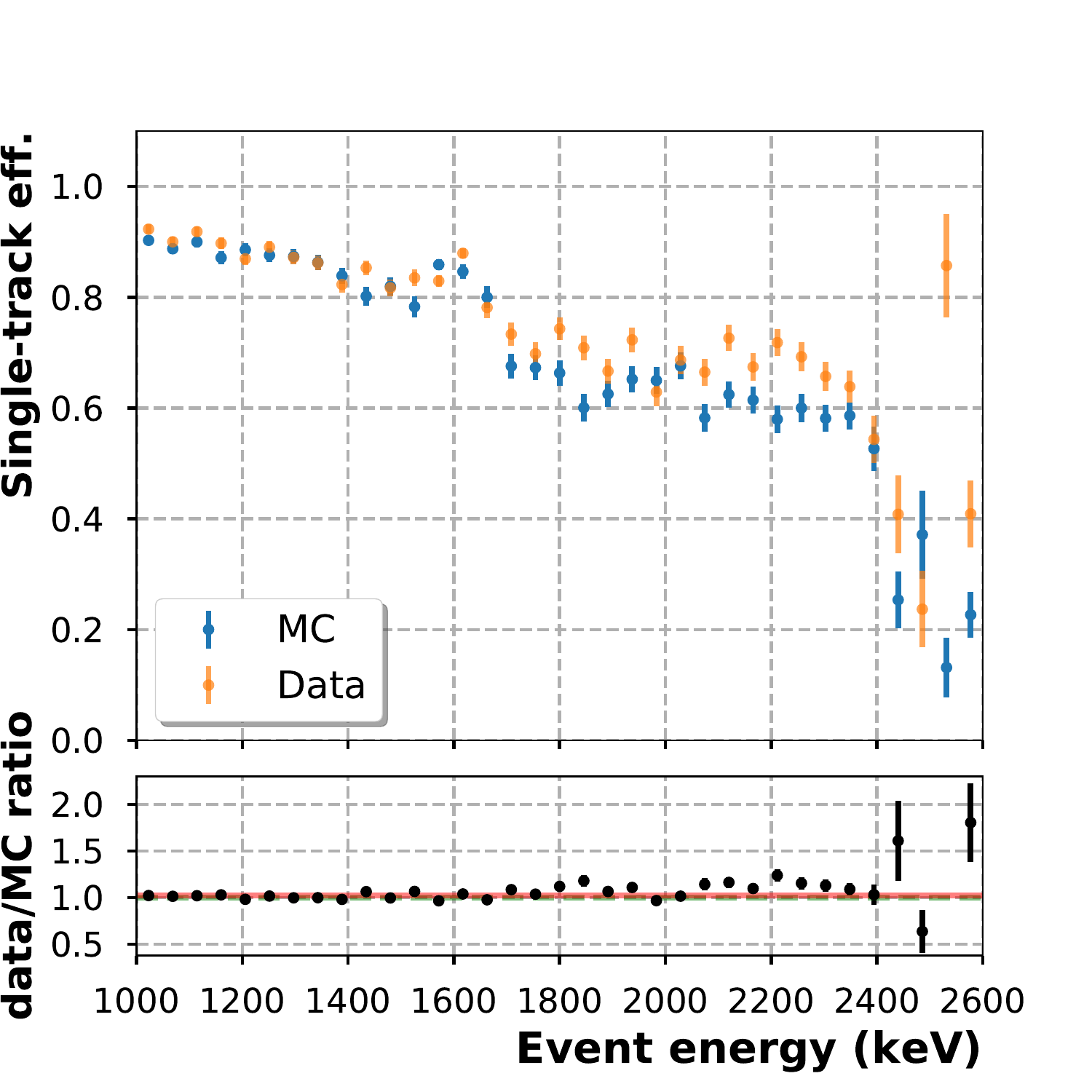}
    \includegraphics[scale=0.4]{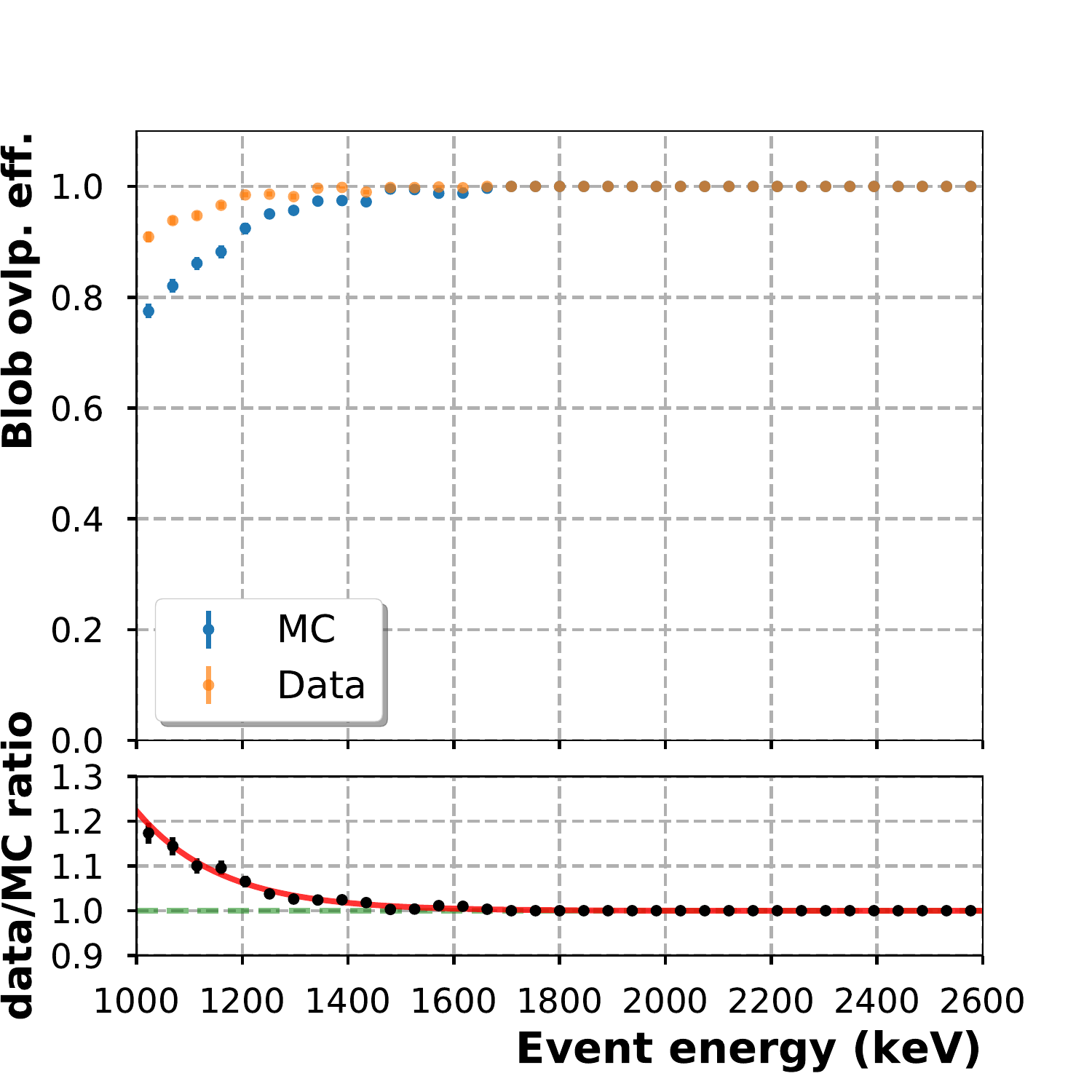}
    \caption{Efficiency of single track (left) and blob overlap (right) selections, as a function of the event energy. Results for data and MC are displayed with orange and blue dots, respectively. The lower panels show the ratio between data and MC, fitted to a horizontal line and to an exponential plus a constant term, respectively.}
    \label{fig:st-ovlp_eff}
  \end{center}
\end{figure}

As done in Sec.~\ref{sec:fiducialsel}, the efficiency of the selection cuts is evaluated with \Th{232} calibration data and the corresponding \Tl{208} MC. The selection efficiency of the single track and blob overlap cuts is shown in Fig.~\ref{fig:st-ovlp_eff}, as a function of the event energy. The integrated efficiency of the former is (89.8$\pm$0.2)\% in data and (87.4$\pm$0.3)\% in MC. The energy dependence reflects the fact that the mean track length of electrons increases with energy, as does the probability for an event to be wrongly reconstructed as a multi-track one. The integrated efficiency of the blob overlap cut is (98.3$\pm$0.1)\% and (94.5$\pm$0.2)\% in data and MC, respectively. In this case, the efficiency increases with energy (i.e., track length), until it reaches $\sim$100\% around 1300~keV. While the MC reproduces well the efficiency of the single track requirement, a significant deviation is observed for the blob overlap cut below $\sim$1300~keV. This disagreement comes from a difference in the data and MC reconstructed track lengths, with MC tracks being shorter. The origin of this discrepancy is under investigation. However, its impact can be accurately described, and thereby corrected for, via an exponential plus a constant term as shown in the lower right panel of Fig.~\ref{fig:st-ovlp_eff}.

\begin{figure}
  \begin{center}
    \hspace*{-0.25cm} 
        \includegraphics[scale=0.4]{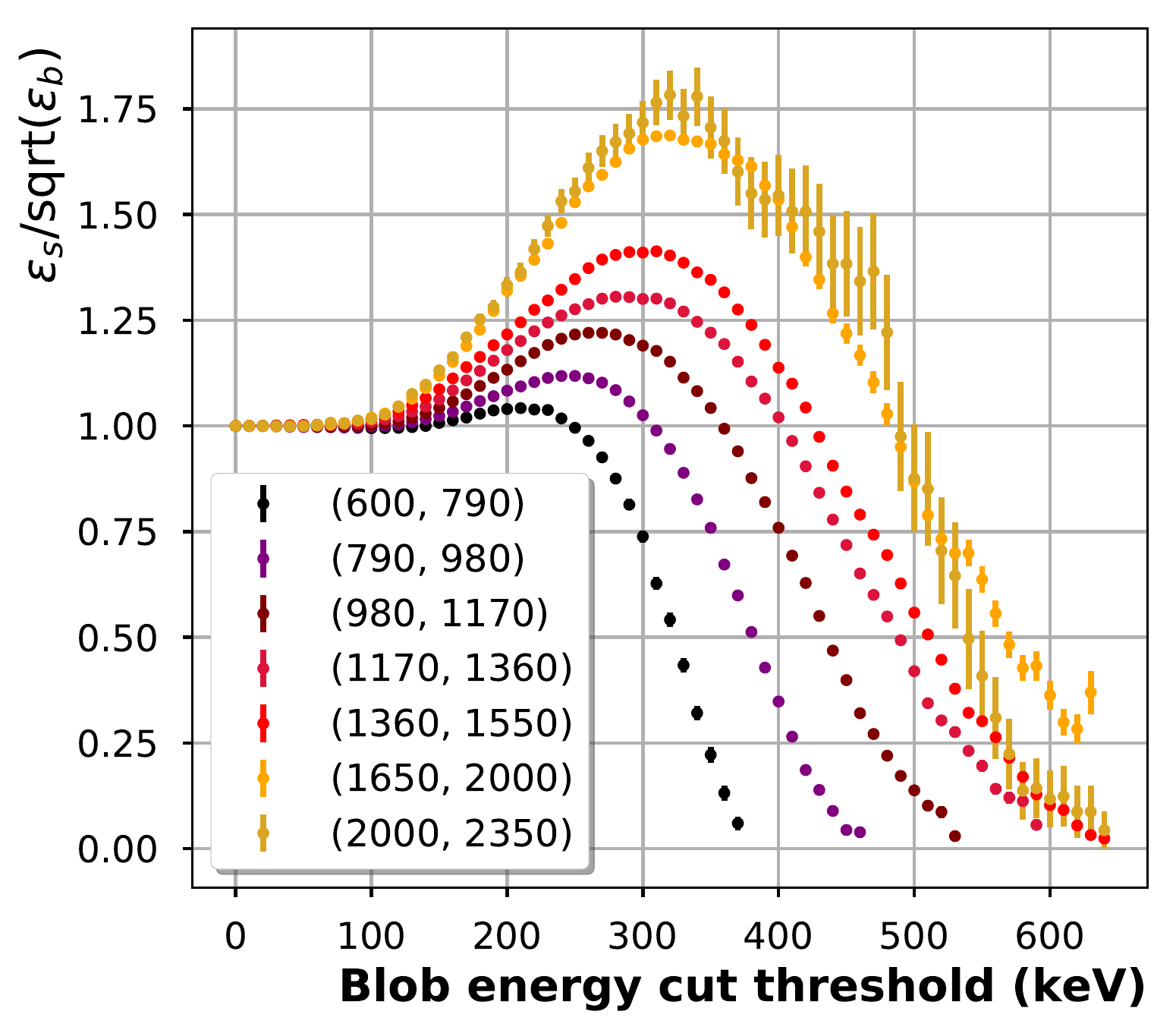}
     \hspace*{1cm} 
        \includegraphics[scale=0.4]{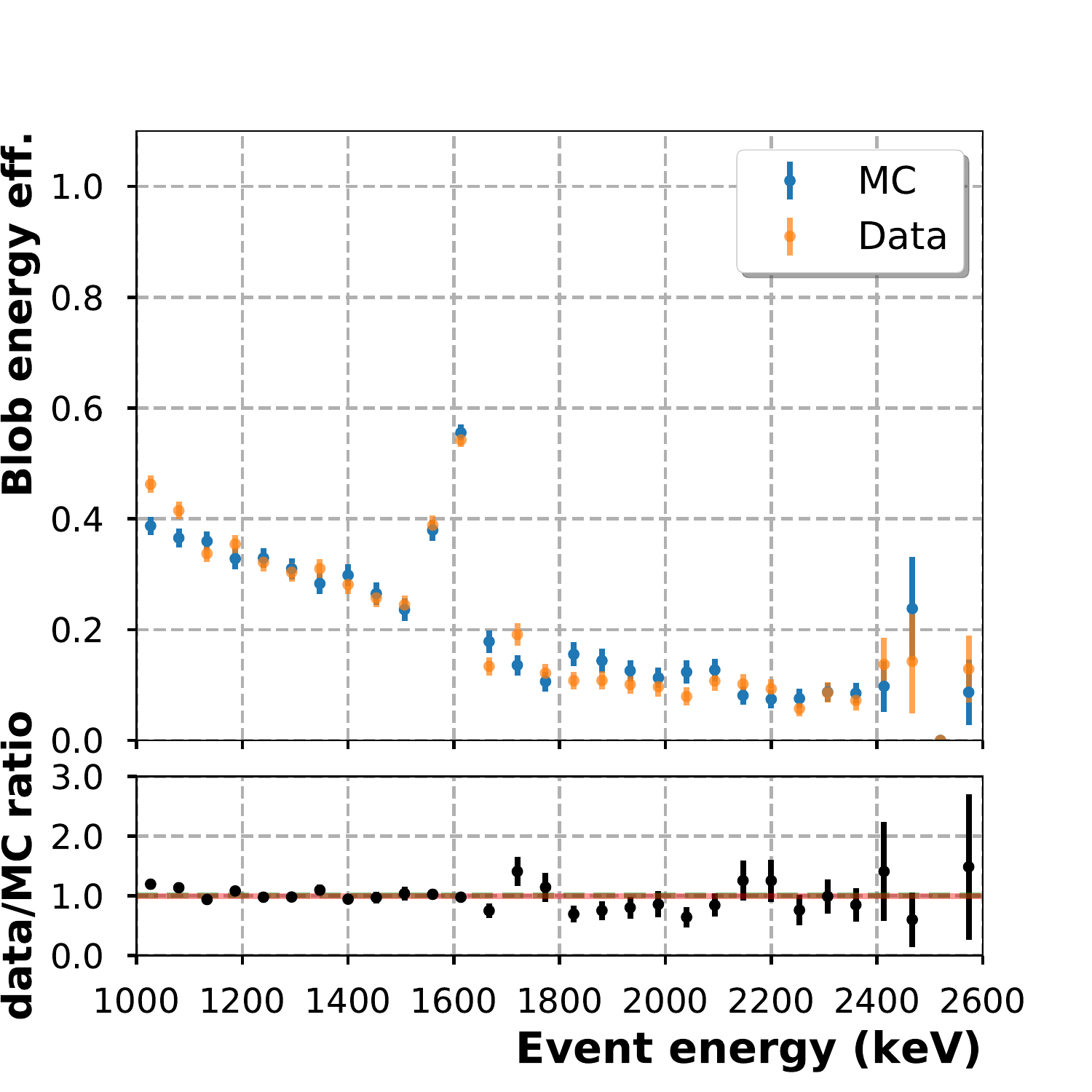}
    \caption{Left: optimization of the blob energy cut (figure of merit) for different track energy ranges. Right: efficiency of the blob energy selection, as a function of the event energy. Results for data and MC are displayed with orange and blue dots, respectively. The lower panel shows the ratio between data and MC fitted to a horizontal line.}
    \label{fig:roc_fom}
  \end{center}
\end{figure}

The energy threshold for the blob energy selection is determined as a function of the track energy by maximizing the figure of merit $\frac{\varepsilon_{signal}}{\sqrt{\varepsilon_{bkg.}}}$, where $\varepsilon_{signal}$ and $\varepsilon_{bkg.}$ are the efficiencies for \bbtwonu signal and background events. A \Xe{136} \bbtwonu MC sample (10$^{6}$ events) is used to estimate $\varepsilon_{signal}$, while the background MC for Run-IVc is used to derive $\varepsilon_{bkg.}$. The values of the figure of merit for different track energy ranges is shown in the left panel of Fig.~\ref{fig:roc_fom}, as a function of the blob energy cut. In this case, the blob energy in MC is corrected by a factor (12.2$\pm$0.6)\% with respect to the data, to account for the difference in the track lengths. The MC optimization in the left panel of Fig.~\ref{fig:roc_fom} results in a higher blob energy cut threshold as the track energy increases. The optimal thresholds for each energy range are fitted to an exponential distribution, which is used as a parametrization to obtain the blob energy cut threshold for each track energy. The corresponding efficiency of the blob cut when applied to data and MC is shown in the right panel of Fig.~\ref{fig:roc_fom}. The integrated efficiencies are (29.3$\pm$0.4)\% and (28.1$\pm$0.4)\% for data and MC, respectively. As the energy at the start-point of a single-electron track decreases as the total electron energy increases, while the blob energy cut threshold increases, the decreasing trend of the efficiency is expected and well reproduced by the MC. The MC reproduces well also the sharp efficiency increase seen in data near the 1592~keV \Tl{208} double-escape peak, consisting of genuine double-electron events.

\section{Backgrounds in double beta decay searches}
\label{sec:betabetasearch}

\begin{figure}
  \begin{center}
    \includegraphics[scale=0.37]{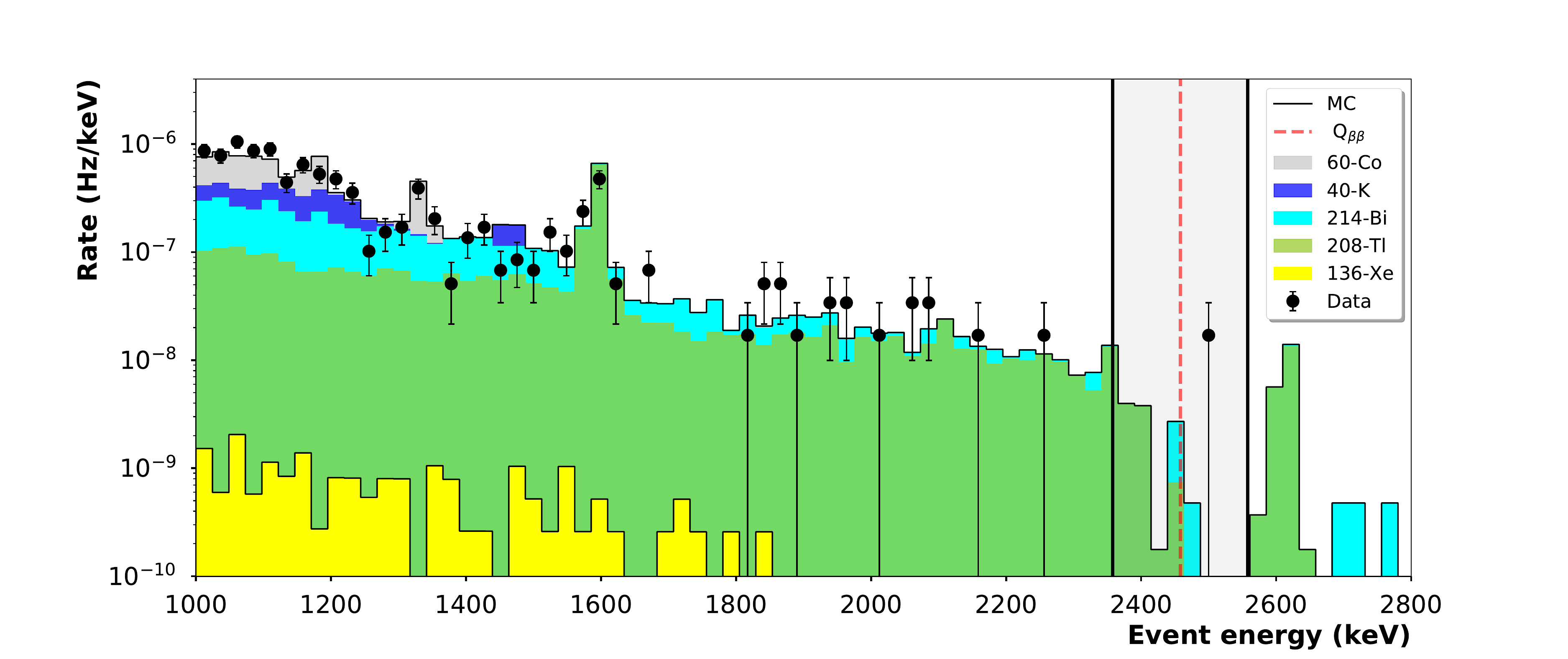}
    \caption{Energy spectrum after topological cuts are applied to background MC and Run-IVc data. The different isotope contributions that form the MC stacked histogram are shown with different colors. The light grey band shows the Q$_{\bb}\pm$100~keV window considered to compare the data to the MC expectation around  Q$_{\bb}$.}
    \label{fig:topology_spectrum}
  \end{center}
\end{figure}

In order to evaluate the backgrounds in \bb analyses, the double-electron selection cuts described in Sec.~\ref{sec:topology} are applied to the Run-IVc fiducial data and MC background samples. In this case, the background model has been rescaled by the normalization factors obtained from the fit in Sec.~\ref{sec:backgroundfit} and the minor differences in the selection efficiencies found between data and MC. The corresponding background rates are shown in Tab.~\ref{tab:topology_rates}. The systematic uncertainties in the MC expectations are derived from the corrections applied to account for the different selection efficiencies in data and MC. The consistency between the rates in data and MC ensures the validity of the background model also after the topological selection. The background rejection factor due to the double-electron selection, with respect to the fiducial sample, is found to be about 3.4 for $E>1000$~keV. The background spectra after topological cuts are shown Fig.~\ref{fig:topology_spectrum}, illustrating a good agreement between data and MC despite the limited statistics. 

\begin{table}[!htb]
\caption{\label{tab:topology_rates} Background rates in Run-IVc data and MC, for E$>$1000 keV and after subsequent topological cuts.}
\begin{center}
\begin{tabular}{ccccc}
\hline
Selection cut & Data rate (mHz)  & MC rate (mHz)\\ \hline
Single track & 0.743$\pm$0.018 & 0.751$\pm$0.002$_{\rm stat}\pm$0.004$_{\rm syst}$\\
Blob overlap & 0.721$\pm$0.017 & 0.721$\pm$0.002$_{\rm stat}\pm$0.018$_{\rm syst}$\\
Blob energy & 0.248$\pm$0.010  & 0.246$\pm$0.001$_{\rm stat}\pm$0.008$_{\rm syst}$\\
\hline 
\end{tabular}
\end{center}
\end{table}

\begin{figure}
  \begin{center}
      \includegraphics[scale=0.30]{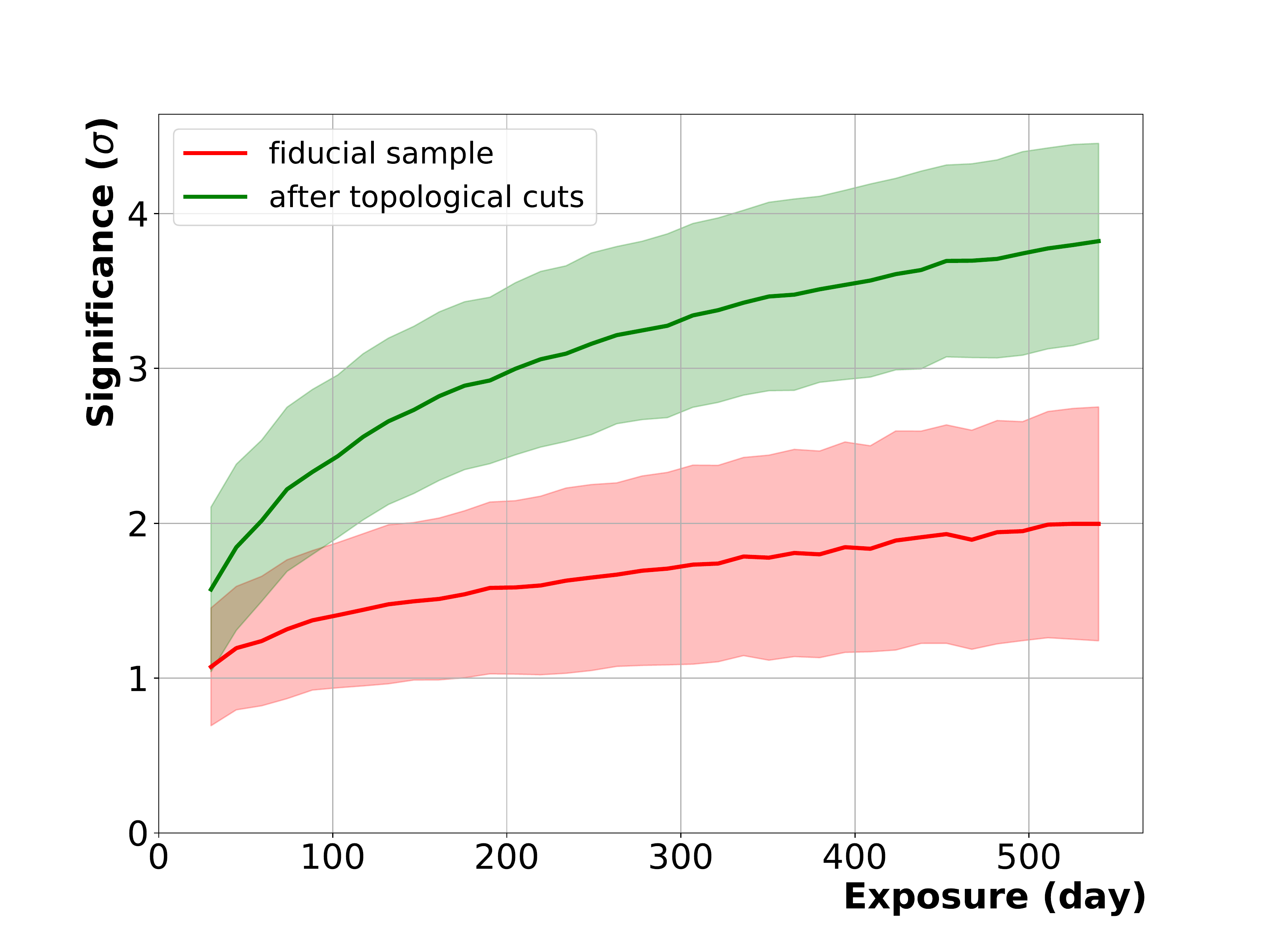}
    \caption{Sensitivity to the \Xe{136} \bbtwonu signal in NEXT-White as a function of exposure, after fiducial cuts only (red) and after all topological cuts (green). Both the average sensitivities as well as the sensitivity RMS spreads are shown.}
    \label{fig:sensitivity}
  \end{center}
\end{figure}

Once validated, the background model after \bb cuts is used to estimate the sensitivity of the NEXT-White detector to the \bbtwonu half-life of \Xe{136}. The sensitivity assumes the same conditions of the ongoing Run-V operations, with enriched xenon (90.9\% \Xe{136} isotopic abundance) at 10.1~bar pressure. The \Xe{136} \bbtwonu signal half-life is taken to be $\halflife = 2.165\times 10^{21}$~yr, following \cite{Albert:2013gpz}. The background expectations are taken to be the ones discussed in Sec.~\ref{sec:backgroundfit}, where the nominal background model expectations have been rescaled to match Run-IVc data yields. For each exposure considered, 1000 toy experiments are generated according to these signal and background expectations. Then, for every toy experiment, an extended maximum likelihood 5-parameter fit to the energy spectrum is performed, where the 5 components are the four background normalization factors (\Co{60}, \K{40}, \Bi{214} and \Tl{208}) plus the signal \Xe{136} \bbtwonu normalization factor.  From the fit, the value and the error of the signal normalization fit parameter is extracted, and a signal sensitivity is computed as the value/error ratio. Since the Run-IVc data offers a direct measurement of the backgrounds, the fits are constrained by the measurement of the \Co{60}, \K{40}, \Bi{214} and \Tl{208} contributions provided in this work, taking into account also the correlations among isotopes obtained in Sec.~\ref{sec:backgroundmodel}. The fits are repeated both for a fiducial sample and for a sample after topological cuts, for each exposure value, and for each toy experiment. The mean sensitivity averaged over the toy experiments is shown with thick solid lines in Fig.~\ref{fig:sensitivity} as a function of exposure. The bands in the figure give the sensitivity RMS spreads obtained from the 1000 toy experiments. According to these results, a (3.5$\pm$0.6)$\sigma$ measurement of the \bbtwonu half-life can be achieved in NEXT-White after 1 year, applying topological cuts. The sensitivity deteriorates significantly if only fiducial cuts are applied. The operation of NEXT-White with enriched xenon will continue until summer 2020, when the installation of NEXT-100 starts. Accounting for the current DAQ dead-time in Run-V, a total live time of about one year is expected.

\begin{figure}
  \begin{center}
    \includegraphics[scale=0.39]{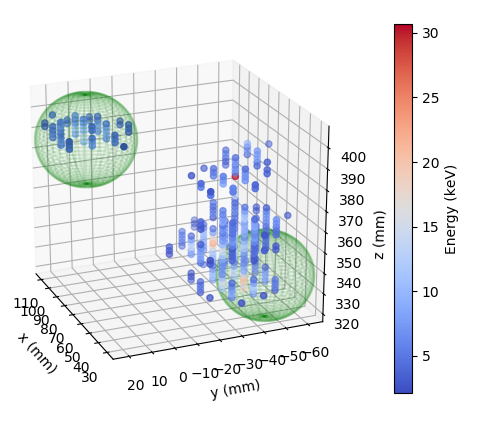}
    \includegraphics[scale=0.39]{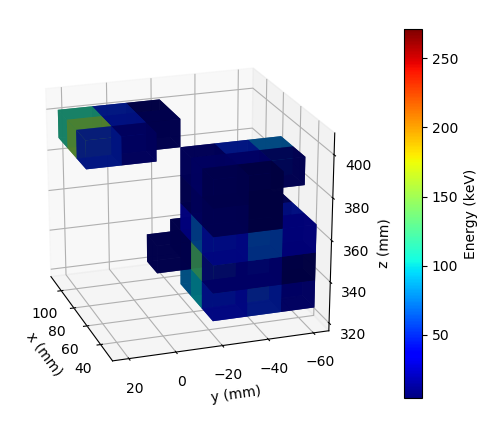}
    \caption{Display of the Run-IVc event that passes the loose \bbnonu selection cuts. Left panel: energy-corrected hits. Blobs are represented as green spheres. Right: voxelized track of the same event; all voxels are connected.}
    \label{fig:display}
  \end{center}
\end{figure}

The \bbnonu backgrounds are also evaluated in an energy window around the Q$_{\bb}$ of \Xe{136} (2458 keV). A {\it loose \bbnonu selection} is defined as the topological selection plus a Q$_{\bb}\pm$100~keV event energy requirement. Although this energy region is not representative of the $\sim$1\% FWHM energy resolution of the detector, it provides a statistically meaningful data/MC comparison using only 37.86~days of Run-IVc data and avoids the 2615~keV \Tl{208} photo-peak. This is the area shown by the light grey band in Fig.~\ref{fig:topology_spectrum}. One event is found to pass the loose \bbnonu cuts in the entire Run-IVc period, in agreement with a MC expectation of (0.75$\pm$0.12$_{\rm stat}\pm$0.02$_{\rm syst}$) events. This provides a validation of the background model also in the \bbnonu region of interest. Out of the background MC events passing the loose \bbnonu cuts, 81\% (19\%) correspond to \Tl{208} (\Bi{214}) decays. The \Tl{208} events come mostly from the anode (60\%) and the cathode (28\%) regions. In the case of the \Bi{214} events, 80\% originate at the anode and 20\% at the cathode regions. Given the good data/MC agreement, the model can be used to estimate the background rejection near \Qbb. With respect to the fiducial sample, an average rejection factor of 16.8$\pm$2.2 due to the \bb selection alone is obtained for the entire (\Tl{208} plus \Bi{214}) high-energy background sample. Concerning the single Run-IVc event passing the cuts, a visual scan has been performed. Fig.~\ref{fig:display} shows a 3D display of this event in terms of SiPM hits and energy blobs (left), and in terms of the corresponding voxels built for the track reconstruction (right). From the comparison between the two panels, it can be concluded that the event consists of two tracks wrongly reconstructed as a single one due to the 15~mm size of the voxels. This indicates that improvements in conventional reconstruction algorithms (see for example \cite{Simon:2017pck}) should lead to better background suppression. In addition, topological reconstruction based on Deep Neural Networks can provide further background reduction \cite{Renner:2016trj}.

\section{Conclusions}
\label{sec:conclusions}

The search for neutrinoless double beta decay (\bbnonu) with a sensitivity exceeding the current half-life upper limit of $\sim$10$^{26}$y \cite{KamLAND-Zen:2016pfg,Agostini:2018tnm} implies a significant experimental challenge. Given the typical 2--3~MeV Q-values of the most promising \bb emitters, natural radioactivity is one of the main backgrounds. The NEXT Collaboration is conducting an experimental program based on electroluminescent high-pressure xenon gas TPCs for \bbnonu searches. The NEXT-White detector is the first large-scale and radiopure implementation of the NEXT technology. The detector holds about 5 kg of xenon at $\sim$10 bar and has been operated at the LSC since 2016. NEXT-White has collected over 3 months of data (Run-IV) with depleted xenon ($<$3\% of \Xe{136}) with the goal of measuring the backgrounds levels. 

The Run-IV background data are divided into three periods, accounting for different operation conditions (Run-IVa, Run-IVb and Run-IVc). During the first period, the radon abatement system (RAS) was not yet available, while radon-free air was delivered continuously by the RAS during the second and third periods. The time correlation of the airborne radon in the LSC with the daily background measurements during Run-IVa has provided an estimation of the background rate above 600 keV in absence of external \Rn{222} (3.97$\pm$0.46~mHz). The consistency with the background measurement in Run-IVb (3.90$\pm$0.06~mHz) proves that the RAS allows for virtually airborne radon-free operations of NEXT-White. The data taken in Run-IVc, with the inner lead castle installed, have been used to measure the background in the same conditions as in Run-V (\Xe{136}-enriched operation), devoted to the measurement of the \bbtwonu half-life. The fiducial background rate is found to be (2.78$\pm$0.03$_{\rm stat}\pm$0.03$_{\rm syst}$)~mHz above 600 keV.

The Run-IVc data have been confronted by a background model considering four radioactive isotopes (\Co{60}, \K{40}, \Bi{214} and \Tl{208}) and 22 detector volumes. The model is built with a GEANT-4 simulation relying on the activity screening of 44 detector materials \cite{Alvarez:2012as,Alvarez:2014kvs,Cebrian:2017jzb} and the internal Rn expectation from \cite{Novella:2018ewv}. The comparison with the background measurement is performed for $E>1000$~keV to neglect possible low-energy contributions not accounted for in the model. The predicted background rate is (0.489$\pm$0.002$_{\rm stat}\pm$0.004$_{\rm syst}$)~mHz while it is found to be (0.84$\pm$0.02)~mHz in data, yielding a ratio of 1.72$\pm$0.04. A fit of the energy and z distributions of the data to the background model provides a measurement of the specific rate of each isotope contribution: R(\Co{60})=(0.23$\pm$0.02)~mHz, R(\K{40})=(0.13$\pm$0.02)~mHz,  R(\Bi{214})=(0.22$\pm$0.04)~mHz, and R(\Tl{208})=(0.27$\pm$0.02)~mHz. The sensitivity of the fit to the spatial origin of the backgrounds also indicates that most of the excess with respect to the model comes from the anode region. 

In order to evaluate the corresponding background in the \bb searches, a set of topological cuts have been applied, requiring the events to be reconstructed as single-track, double-electron events. The background rate after the topological selection is (0.248$\pm$0.010)~mHz and (0.244$\pm$0.001$_{\rm stat}\pm$0.008$_{\rm syst}$)~mHz for the data and the MC expectation, respectively. In this case, the background model contributions have been scaled according to the background fit results. According to the background model, a background reduction of $\sim$3.4 for $E>1000$~keV is achieved by means of the topological information of the events. The best-fit background model has been used to estimate the sensitivity of NEXT-White to the \bbtwonu half-life, which is found to be (3.5$\pm$0.6)$\sigma$ after one year of data taking. Concerning the search for \bbnonu decay, the expected background in a 200~keV window around the Q$_{bb}$ of \Xe{136} is 0.75$\pm$0.12$_{\rm stat}\pm$0.02$_{\rm syst}$ in 37.9 days, while 1 event is observed in the Run-IVc data. Thus, the background model tuned using lower-energy data ($E>1000$~keV) is also validated in this higher-energy \bbnonu range. For this energy window, the topological selection yields a background reduction of 16.8$\pm$2.2, the remaining events being dominated by the contribution from the anode region. 

Overall, the results derived from NEXT-White Run-IV data validate the background assumptions used to estimate the physics case of the NEXT-100 experiment \cite{Martin-Albo:2015rhw} and provide essential inputs to improve the detector design. It has been shown that the contribution from airborne \Rn{222} to the \bbnonu backgrounds will be negligible, thanks to the RAS. Concerning the radiogenic backgrounds from the detector materials, the reliability of the model has been confirmed with Run-IVc data, in particular in a 200~keV window around Q$_{bb}$. Assuming the same level of radio-impurities in the detector materials, the \bbnonu background index in NEXT-100 is expected to decrease with respect to NEXT-White from geometrical considerations alone, although the exact background scaling will depend on the precise background origin. Concerning NEXT-100 design and installation, NEXT-White background data have identified the anode (tracking plane) region as the detector area where improvements with respect to NEXT-White could be particularly beneficial in terms of achievable background levels.

\acknowledgments
The NEXT Collaboration acknowledges support from the following agencies and institutions: the European Research Council (ERC) under the Advanced Grant 339787-NEXT; the European Union’s Framework Programme for Research and Innovation Horizon 2020 (2014-2020) under the Marie Skłodowska-Curie Grant Agreements No. 674896, 690575 and 740055; the Ministerio de Econom\'ia y Competitividad and the Ministerio de Ciencia, Innovaci\'on y Universidades of Spain under grants FIS2014-53371-C04, RTI2018-095979, the Severo Ochoa Program SEV-2014-0398 and the Mar\'ia de Maetzu Program MDM-2016-0692; the GVA of Spain under grants PROMETEO/2016/120 and SEJI/2017/011; the Portuguese FCT under project PTDC/FIS-NUC/2525/2014, under project UID/FIS/04559/2013 to fund the activities of LIBPhys, and under grants PD/BD/105921/2014, SFRH/BPD/109180/2015 and SFRH/BPD/76842/2011; the U.S.\ Department of Energy under contracts number DE-AC02-06CH11357 (Argonne National Laboratory), DE-AC02-07CH11359 (Fermi National Accelerator Laboratory), DE-FG02-13ER42020 (Texas A\&M) and DE-SC0019223 / DE-SC0019054 (University of Texas at Arlington); and the University of Texas at Arlington. DGD acknowledges Ramon y Cajal program (Spain) under contract number RYC-2015-18820. We also warmly acknowledge the Laboratori Nazionali del Gran Sasso (LNGS) and the Dark Side collaboration for their help with TPB coating of various parts of the NEXT-White TPC. Finally, we are grateful to the Laboratorio Subterr\'aneo de Canfranc for hosting and supporting the NEXT experiment.

\bibliographystyle{JHEP}
\bibliography{biblio}

\end{document}